# ISTQB Certifications Under the Lens: Their Contributions to the Software-Testing Profession; and AI-assisted Synthesis of Practitioners' Endorsements and Criticisms

| Vəhid Gəruslu[1] | Alper Buğra Keleş | Sevde Değirmenci |
|---|---|---|
| Queen's University Belfast, UK<br>v.garousi@qub.ac.uk<br><br>Azerbaijan Technical University, Azerbaijan<br>vahid.garousi@aztu.edu.az | Testinium A.Ş., Türkiye<br>alper.keles@testinium.com | Q Investment Bank, Türkiye<br>s.degirmenci@qyatirimbankasi.com.tr |

**Abstract:**

***Context:*** The International Software Testing Qualifications Board (ISTQB) certification dominates global software testing practice, with 1.2+ million certifications issued across 130+ countries. Yet it remains contested: practitioners value it for career advancement and shared terminology, while others criticize it as overly theoretical and somewhat disconnected from real-world testing practices.

***Objective:*** This study investigates the perceived value and critique of ISTQB certifications. While the certifications aim to standardize the software testing body of knowledge, debates persist about their practical relevance and impact. Our objective was to systematically capture practitioner perspectives and assess the precision of endorsements and fairness of criticisms through expert review.

***Method:*** We conducted an AI-assisted Multivocal Literature Review (MLR), combining academic and grey literature to synthesize practitioner endorsements (RQ1) and criticisms (RQ2). ChatGPT's deep research capability was employed under continuous human oversight, with QA strategies ensuring transparency and reliability. As another analysis, we asked a panel of four independent experts to evaluate the precision of endorsements and fairness of criticisms.

***Results:*** Practitioner endorsements emphasized career benefits, improved communication, and a shared vocabulary as the main values of ISTQB certifications. Criticisms focused on excessive theoretical content, limited relevance in agile and automation-intensive contexts, and weak support for real testing skills. Expert review confirmed that while many endorsements were precise, several criticisms reflected broader tensions in the discipline, including contrasting schools of thought in testing practice.

***Conclusions:*** ISTQB certifications provide recognizable career and communication value but remain contested in terms of practical utility. By triangulating practitioner voices with expert validation, this study delivers an evidence-based reflection on the strengths and weaknesses of ISTQB in shaping the software testing body of knowledge. The AI-assisted MLR also demonstrates how GenAI tools can support systematic evidence synthesis when coupled with rigorous human oversight.

[1] Alternate name spelling: Vahid Garousi

# 1 INTRODUCTION

Software testing qualifications play a vital role in shaping the competence, credibility, and career progression of testers. In the global software industry, professionals arrive at testing roles through diverse pathways. Some enter the field via formal academic degrees—such as software engineering, computer science, or dedicated testing programs—while others transition from adjacent roles or become self-taught through practical exposure [1-3]. In recent decades, a growing number of practitioners have also pursued structured qualifications to formalize and validate their knowledge, particularly through globally recognized certification schemes [4, 5].

Among these, the certification program developed by the International Software Testing Qualifications Board (ISTQB) stands as the most widely adopted and influential. Founded in Edinburgh, Scotland in 2002, the ISTQB is a non-profit association legally registered in Belgium. Since its inception in 2002, and as of May 2026, ISTQB has administered over 1.5 million exams and has issued more than 1.2 million certifications globally, across more than 130 countries. Its tiered qualification structure—comprising Foundation, Advanced, Specialist, and Expert levels—has become a key reference point for hiring, training, and evaluating testers in both private and public sectors.

The extensive adoption of ISTQB on the worldwide scale (noting its 1.2 million certifications) not only underscores its impact on hiring, training, and professional development but also motivates the need to critically examine and understand how practitioners perceive the value and limitations of a certification scheme with such global influence.

Despite its broad reach and institutional maturity, ISTQB certifications have also attracted considerable critique. Observers have questioned the real-world relevance of the exams, the overemphasis on memorization, the limited assessment of practical skills, and the commercialization of training ecosystems around the certifications. At the same time, others regard ISTQB as an essential entry point—providing structure, a common language, and baseline expectations for professional testers, particularly in regions with limited access to formal testing education.

Studying the ISTQB certification ecosystem offers important benefits for both researchers and practitioners [6]. For researchers, it presents an opportunity to better understand how formal qualification schemes shape the evolution of professional roles, influence industry norms, and define what constitutes knowledge in software testing. It also opens up lines of inquiry into how standardization efforts interact with real-world practices and learning pathways. For practitioners—including testers, hiring managers, and training coordinators—this research provides insights into the actual utility, reception, and perceived shortcomings of ISTQB certifications. Such knowledge can guide more informed decisions about certification uptake, hiring criteria, and training investments, particularly in environments where certifications are assumed to serve as proxies for skill or professional maturity.

One of ISTQB's notable contributions is the systematic structuring of testing knowledge into role-based syllabuses with clearly defined learning objectives. As of 2026, the ISTQB portfolio includes more than 22 syllabuses covering core testing topics and specialized areas such as automation, AI, security, game testing, and performance engineering. These documents, developed collaboratively by international working groups and reviewed through version-controlled processes, help ensure a level of consistency and rigor. While extensive literature on software testing has existed for decades—spanning academic books, research papers, and practitioner insights—ISTQB has played a unique role in distilling, organizing, and operationalizing this body of knowledge into a form that is actionable, teachable, and certifiable for professionals across the globe. For ease of reference, we refer to this structured and role-aligned knowledge base as the Software Testing Body of Knowledge (STBoK), recognizing that the term itself is not yet formally established.

This paper offers a critical and constructive reflection on ISTQB's role in the software testing profession. It pursues three main objectives, which are also the main contributions of this paper:

1. A pragmatic and consolidated non-trivial review of the ISTQB certifications and their contributions to the software-testing profession, e.g., discussing their contributions to the STBoK, their technical breadth and evolution, their utility for benchmarking of testers, hiring and career planning (Section 3)
2. A systematic synthesis of perceptions of value, benefits and critique of ISTQB certifications by practitioners as shared online, i.e., in the grey literature (Section 5)
3. Analyzing the extent to which the practitioner endorsements and criticisms of ISTQB certifications are considered precise and fair, when evaluated by a group of independent experts (Section 6)
4. A documented process, plan and execution details for conducting AI-assisted Multivocal Literature Reviews (MLR) – presented in Section 4.2 and also in a supplementary document [7].

The structure of this paper is as follows. Section 2 provides a background on software testing education, qualification pathways, and certification schemes, including ISTQB. Section 3 provides a review of ISTQB and its contributions to the software testing body of knowledge. Section 4 details our research design and methodology. Section 5 presents the findings from the MLR: the synthesized set of endorsements and criticisms of ISTQB certifications. Critical reviews on precision of endorsements, and fairness of criticisms, done by a group of independent experts, are presented in Section 6. Discussions, implications and limitations are discussed in Section 7. Section 8 concludes and suggests directions for future research.

## 2 BACKGROUND AND RELATED WORK

This section is intentionally broad and detailed, as it aims at providing the essential context for understanding the significance, reception, and impact of ISTQB certifications. Our aim is to build a solid background by covering multiple dimensions of software testing education and certification.

We begin by situating ISTQB within the broader landscape of software engineering (SE) and software testing education and training (Section 2.1), followed by an overview of different career pathways into the testing profession (Section 2.2). We then examine the respective roles of universities and industry in preparing testers (Section 2.3), present a historical and structural review of ISTQB certifications (Section 2.4), and offer a comparative look at other certification schemes (Section 2.5).

We then introduce the notion of a STBoK (Section 2.6), and synthesize related academic and grey literature (Section 2.7). Finally, we reflect on the related experience of the authors (Section 2.8), and how this expertise informs the perspectives and evidence presented in this paper.

### 2.1 The landscape of software-testing education and training

Most CS/SE programs in universities offer software-testing courses as part of their curricula, while industry provides training through certification schemes and commercial providers. Together, these sources form a diverse ecosystem of education and training in software testing.

There is a very substantial number of software testers worldwide. While there are no official statistics, rough estimates place the number between 5–9 million practitioners [8, 9]. Training and educating this large existing workforce, as well as the new generations of testers entering the field, is therefore a major undertaking with significant implications for both academia and industry.

A 2020 systematic literature mapping [2] showed steady growth in research on how testing is taught, with methods ranging from lectures to project-based learning and game-based exercises. Earlier surveys [1] noted that testing courses exist in most curricula but often vary in depth and practical emphasis. Subsequent work introduced laboratory courseware [10], use of industrial-strength systems under test [11], and incorporation of real-world industrial projects [8], all showing that students benefit most when education integrates hands-on experience.

To train the next generation of software testes, more recent analyses highlight the importance of closer cooperation between academia and industry [3, 12]. Effective preparation of test engineers requires curricula that keep pace with industrial needs, which can be achieved through co-designed courses, guest lectures, and student participation in live projects.

Beyond academia, the commercial training market is substantial. Numerous companies offer short courses, bootcamps, and vendor-specific programs. Many are aligned with certification frameworks such as ISTQB, while others target specialized skills like automation or performance testing. This reflects the strong demand for continuous upskilling in a rapidly evolving profession.

Overall, the landscape of software-testing education and training combines academic foundations, certification-based knowledge frameworks, and market-driven training programs. Stronger integration among these dimensions is still needed to balance conceptual understanding with practical competence [1-3, 10].

### 2.2 Different paths to becoming a software tester

This diversity of educational and training routes is visualized in Figure 1, which was published in [3]. This model illustrates the multiple pathways through which individuals enter and grow within the software-testing profession. University graduates may come from SE/CS/IT programs with varying degrees of exposure to testing, or from unrelated fields where testing was not taught as part of the curricula. Others are self-taught engineers who acquire skills through practice and self-learning. At different points in their careers, testers may pursue industrial training, vendor-specific courses, or certifications such as ISTQB.

Many organizations also provide in-house training to build testing capacity tailored to their own contexts. The figure emphasizes the heterogeneous competency profiles of testers, shaped by a combination of conceptual knowledge and practical experience, and highlights why a one-size-fits-all approach to education and certification cannot fully address the needs of such a diverse workforce.

Discussing this diversity is useful and important for the present paper, because it shows why certifications like ISTQB receive both strong endorsements and strong criticisms. For some practitioners, ISTQB aligns well with their educational background and career goals, while for others, especially those trained primarily through hands-on or alternative routes, the certification's focus may feel too theoretical or insufficiently practice-oriented. Recognizing these varied pathways helps contextualize the polarized perceptions analyzed later in this study.

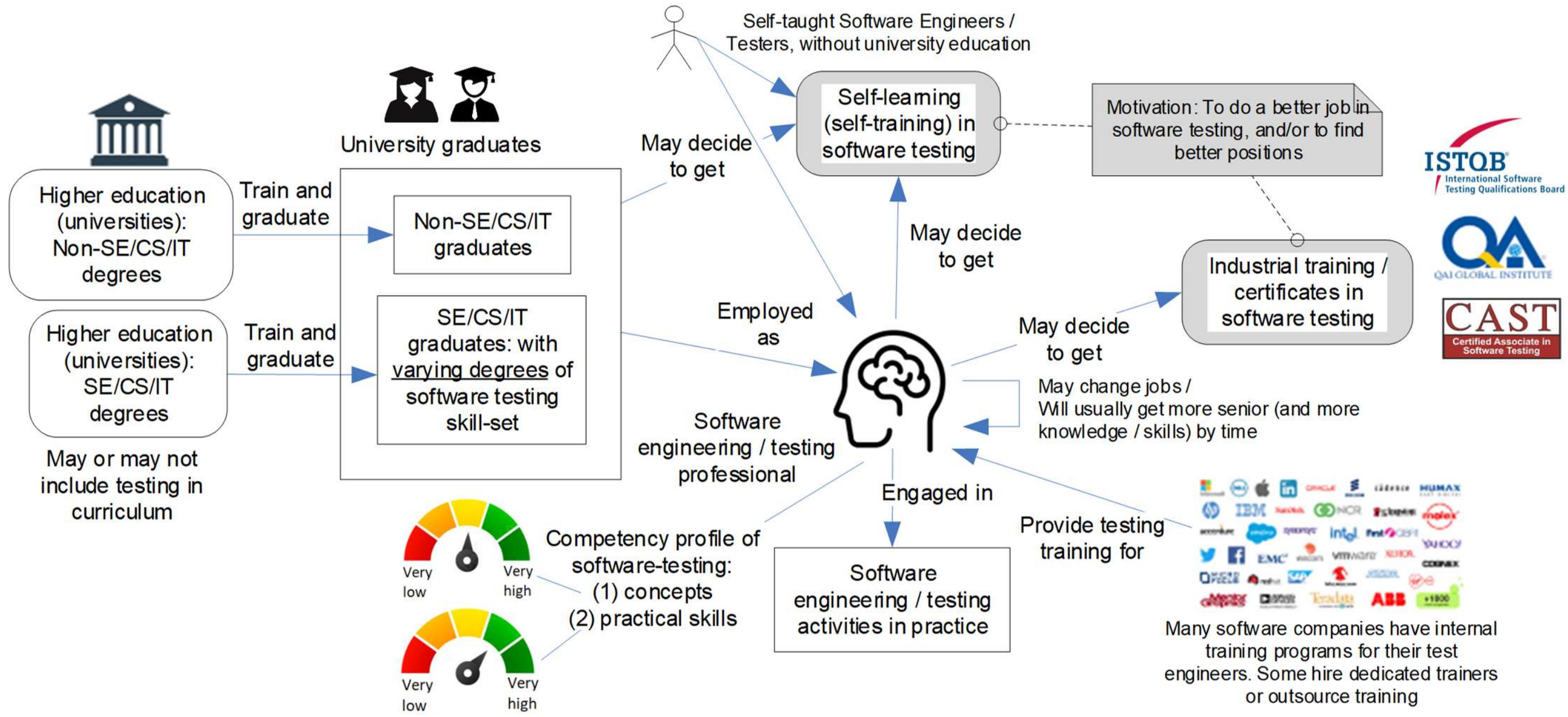

**Figure 1- A model showing the diversity of educational and training routes and the relationship of software-testing education in universities versus training in industry (published in [3])**

## 2.3 A review and history of the ISTQB Certifications

The certification scheme leading to the ISTQB originated in the United Kingdom in 1998, when the British Computer Society (BCS)'s Information Systems Examination Board (ISEB) introduced the first Foundation Level exam and established a software testing commission. This initiative marked the beginning of structured, exam-based validation of testing knowledge. As demand for international standards grew, the ISTQB was founded in 2002. Around the same time, similar initiatives emerged in other countries, such as ASQF[1] in Germany, contributing to ISTQB's global spread.

Since then, ISTQB has expanded rapidly, issuing over a million certificates across multiple levels and specializations. While a detailed pragmatic review is presented in Section 3, this brief historical overview shows how ISTQB built upon early national initiatives to become the dominant global standard in software testing certification.

## 2.4 Other certifications in the area of software testing

While ISTQB has become the dominant global scheme, there are numerous other certifications in software testing offered by international organizations. These alternatives often focus on specific methodologies, tools, or frameworks and provide practitioners with options beyond the ISTQB pathway.

iSQI (International Software Quality Institute, isqi.org) is one of the major players. It offers a variety of test-and-quality certifications covering specialisms like performance testing, usability testing, mobile app testing, security testing, automotive systems, and others, e.g., iSQI Selenium Tester Foundation (CSeT-F).

Other notable certifications include:

---

[1] German term: *Arbeitskreis Software-Qualität und -Fortbildung (ASQF)*. English translation: Working Group for Software Quality and Training

- ICP-ATA (ICAgile Certified Professional in Agile Test Automation), by the International Consortium for Agile (icagile.com) which focuses on designing and implementing test automation strategies within Agile environments, including tool usage and exploratory testing
- Certified Professional Selenium Tester (CPST), by the Global Association for Quality Management (GAQM) (gaqm.org)— specialization in automated testing using Selenium framework
- The Alliance for Qualification (A4Q), AllianceForQualification.com, also offers a number of certifications too, such as A4Q Foundation Level Tester for Appium, A4Q Certified Selenium 4 Tester Foundation, and A4Q Software Development Engineer in Test.

Also, many commercial training providers offer other certification programs, bootcamps, and micro-credentials in specialized domains: e.g. mobile testing, API testing, performance/load/stress testing, security, usability/UX, DevOps test practices. These often align with or compete against ISTQB-aligned courses, depending on region and market demand.

These alternatives serve several roles: they allow practitioners to pick credentials that are more tool- or domain-specific, or that emphasize automation or agile contexts. On the flip side, many of them don't have as wide or consistent global recognition as ISTQB, and their syllabi may vary more in structure, rigor, and update cycles.

## 2.5 Software Testing Body of Knowledge (STBoK)

Software testing has emerged as a recognized field under software engineering since the 1950s, evolving from ad-hoc debugging practices into a discipline with its own methods, techniques, and professional standards. Histories of testing describe its development from manual debugging, through the rise of structured testing techniques in the 1970s and 1980s, to automation, agile testing, and more recently, AI-assisted approaches [14–16].

In parallel, an enormous body of knowledge has accumulated. There are 10,000s of academic papers published on software testing [13-15], covering topics from test design and automation to model-based testing, mutation testing, and AI-driven approaches [14–16]. The field sustains a wide ecosystem of academic and professional conferences, including flagship research venues such as *ICST (International Conference on Software Testing, Verification and Validation)* and *ISSTA (International Symposium on Software Testing and Analysis)*, alongside 1000s of practitioner events each year catalogued on platforms like testingconferences.org. Together, these illustrate both the depth of scientific inquiry and the vibrancy of industrial practice.

Books have also been a major vehicle for consolidating and transferring knowledge. Based on a dataset of published works collected a few years ago [16], more than 350 books on software testing have appeared since the 1970s, spanning practitioner guides, academic textbooks, and specialized volumes (e.g., on test automation, agile testing, performance testing). The classification of these books shows a diversity of emphasis, with some focusing on theoretical foundations, others on methods and techniques, and many on applied industrial practices.

By analogy with the well-established SWEBOK (Software Engineering Body of Knowledge), it is appropriate to speak of a STBoK. The term itself is not yet formally established, nor does an official framework exist, but all the accumulated knowledge—spanning books, research papers, conferences, syllabi, and industrial practices—together already constitutes a de facto STBoK. This broader perspective is important for our study, as ISTQB should be seen as one influential but partial representation within this much richer and more diverse body of knowledge.

## 2.6 Related work and novelty of this work

Several studies [5, 17-21] have directly examined ISTQB certifications or used them as an anchor for broader analyses. We discuss each of those studies next.

Early advocacy is visible in Black's 2008 article [5], which promoted ISTQB as a way to standardize tester competence globally, outlining its perceived benefits and pathways to certification.

Other contributions have emphasized workforce development using ISTQB. For example, the work by Farley [17]discussed the cultivation of software quality engineers, situating ISTQB within broader organizational strategies for competence building, including mentoring and automation practices. This positions certification not as a standalone solution but as part of a wider skill-development ecosystem.

Some studies have aligned ISTQB with established process or standards frameworks. Mahdavifar et al. [19] explored the integration of ISTQB concepts into the U.S. Federal Enterprise Architecture Framework, demonstrating how certification-aligned practices can shape testing processes in government settings. Similarly, Sánchez and Colomo [20] mapped ISTQB syllabi to ISO/IEC/IEEE 29119-2 software testing standard, showing both overlaps and misalignments between certification objectives and formal international standards.

Also, a growing body of work has examined ISTQB in educational contexts. Szatmári et al. [21] analyzed ISTQB-based software testing courses, reporting both advantages (shared terminology, industry recognition) and challenges (risk of overemphasis on theoretical content if not paired with practice). Such findings highlight the complex trade-offs when embedding certification syllabi into academic programs.

Finally, more recent analyses have focused on syllabus evolution. Hamburg and Roman [18] studied the revised ISTQB Advanced Test Analyst syllabus, discussing its incorporation of black-box testing techniques and its orientation toward practitioner applicability.

Taken together, the related work spans advocacy, workforce development, standards alignment, educational integration, and syllabus evolution. This body of literature demonstrates ISTQB's multi-faceted role—as a promotional vehicle, a curricular scaffold, a bridge to international standards, and a living framework adapting to the shifting landscape of software testing.

What distinguishes the present study from prior work, discussed above, is its breadth, methodology, and triangulation of perspectives. Earlier contributions were either promotional, narrow in scope (e.g., mapping ISTQB to a single standard or embedding it in one course), or limited to specific syllabus updates. In contrast, our study undertakes the first systematic, multivocal synthesis of both endorsements and criticisms of ISTQB certifications, drawing simultaneously on academic research, grey literature, and the perspectives of independent experts. By combining these heterogeneous sources into a consolidated evidence base, we provide a more comprehensive and balanced picture of the value and limitations of ISTQB than any previous single-threaded analysis. This methodological integration, supported by AI-assisted review techniques, constitutes the primary novelty of this work.

### 2.7 Related experience of the authors

The perspectives presented in this paper are informed not only by literature but also by the professional and academic experiences of its authors.

The first author has been active in the software testing industry and academia since 1998, working at the intersection of research and practice [22-28]. He wrote his first automated test script in 1998 using IBM Rational Functional Tester and has since continued to combine industrial consultancy with academic research. Since 2006, he has been training software test engineers through university courses, industry workshops, and certification programs. In some of the trainings, he has used the ISTQB curriculum. His consultancy roles have included conducting over 100 hiring interviews for testing positions and mentoring numerous young engineers aspiring to enter the testing profession, often giving them advise to take ISTQB certifications. He also serves on the technical committee of the TMMi Foundation, which some active connections to the ISTQB Association. These experiences provide a unique vantage point on how certifications such as ISTQB are perceived, valued, and critiqued across both academia and industry.

The second author has extensive experience as a QA Manager and Project Leader, having worked in diverse domains such as public safety, finance, e-commerce, and cybersecurity across the USA, Dubai, and Azerbaijan. He has been actively involved in international initiatives such as the ITEA3 TESTOMAT project, focusing on next-generation test automation, and has lectured on advanced software testing at Istanbul Kültür University. His role at Testinium includes providing training, defining testing structures, implementing agile transformations, and aligning test strategies with client needs. His dual role in practice and education allows him to contribute applied insights into how ISTQB certifications are used in industrial projects and training contexts.

The third author brings a complementary background with experience in software testing, quality assurance, and project involvement, as reflected in her professional profile. Her contributions strengthen the paper's grounding in how early-career and practitioner-level testers approach certifications and training, especially in contexts where career entry and progression are strongly shaped by such qualifications.

Together, these experiences reinforce the credibility of this study by anchoring it not only in evidence from the literature but also in firsthand observations and professional practice across academic, industrial, and training contexts.

## 3 A PRAGMATIC AND CONSOLIDATED REVIEW OF THE ISTQB CERTIFICATIONS

This section presents a descriptive synthesis of ISTQB's contributions to the STBoK, based on a detailed review of official syllabus documents, foundational materials published by ISTQB, and relevant academic and grey literature. In addition, the authors draw upon their own extensive professional experience—since the mid-2000s—collaborating with testers who hold ISTQB certifications, mentoring those preparing for certification, and participating in hiring panels where ISTQB qualifications were used as selection criteria. While not based on a formal empirical method, this contextual analysis serves to acknowledge and examine the technical and pedagogical foundations of the ISTQB scheme.

### 3.1 Set of 23 certifications covering different topics of software testing

ISTQB currently offers 23 certifications that cover a broad range of topics in software testing. The certification map (Figure 2, taken from istqb.org/certifications) illustrates this structure, which is organized into Core, Agile, Specialist, and Expert streams. Each stream is designed for different roles, knowledge levels, and areas of specialization, forming a structured set of qualifications.

At the Core level, the *Foundation Level (CTFL)* serves as the mandatory entry point. Candidates must first pass CTFL before progressing further. From this foundation, testers can pursue certifications along multiple streams—much like climbing a ladder—choosing routes according to their interests, job market demands, or organizational requirements. The *Advanced Level* includes options such as *Test Analyst*, *Technical Test Analyst*, and *Test Manager*, which deepen knowledge in analysis, technical aspects, and leadership. At the highest tier, *Expert Level* certifications address organizational test process improvement and advanced management practices.

The Agile stream expands the foundation with certifications such as *Agile Tester (CTFL-AT)*, *Agile Test Leadership at Scale (CT-ATLaS)*, and *Agile Technical Tester (CTAL-ATT)*, reflecting the widespread adoption of agile methods and the demand for testers who can operate effectively in iterative development environments.

The Specialist stream contains a diverse set of certifications targeting niche domains and emerging needs. Examples include *Model-Based Testing (CT-MBT)*, *Test Automation Engineer (CTAL-TAE)*, *Acceptance Testing (CT-AcT)*, *Game Testing (CT-GaMe)*, *Automotive Software Tester (CT-AuT)*, and *AI Testing (CT-AI)*. This breadth illustrates ISTQB's effort to align its syllabi with evolving industry demands, such as safety-critical systems, gaming, and AI-enabled systems.

A notable recent addition is the Certified Tester Testing with Generative AI (CT-GenAI), introduced in 2024. This certification addresses the rising importance of GenAI in the SDLC, equipping testers with knowledge of opportunities, risks, and strategies when applying generative AI tools in testing contexts. It reflects ISTQB's recognition of the disruptive potential of GenAI and the need for testers to acquire structured knowledge in this fast-moving domain.

Finally, the Expert level consolidates advanced knowledge with certifications such as *Improving the Test Process (CTEL-ITP)* and *Test Management (CTEL-TM)*, enabling senior practitioners to specialize in organizational-level improvements and leadership.

Taken together, the 23 certifications cover a spectrum from foundational concepts to specialized areas. This portfolio provides tailored pathways for practitioners, but may also introduces complexity in terms of navigation and selection.

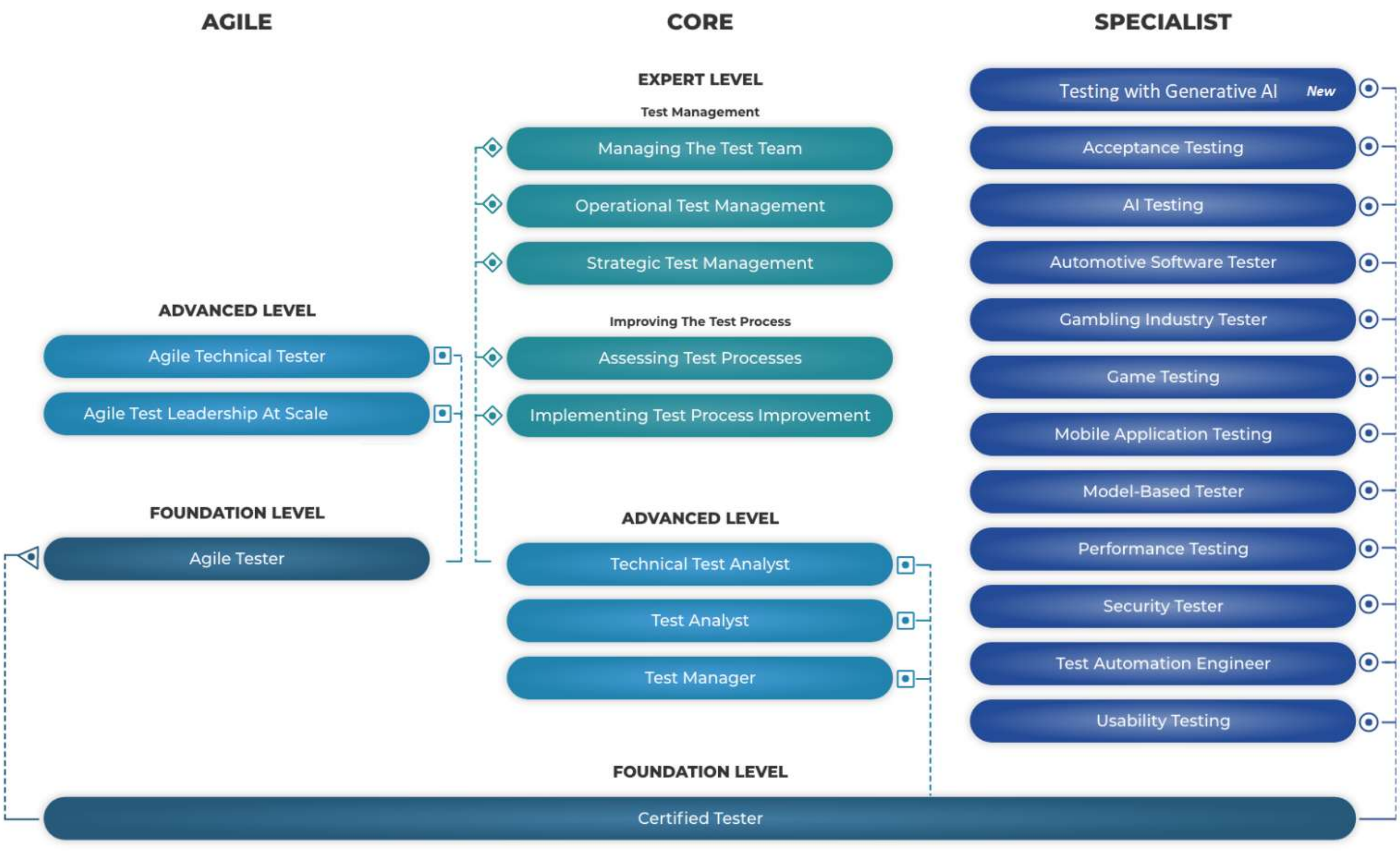

**Figure 2—Map of all ISTQB certifications (istqb.org/certifications)**

## 3.2 Structured, Role-Based Syllabuses

The structured and role-based organization of ISTQB syllabuses is often highlighted as providing clarity and progression for practitioners at different career stages.

The portfolio is arranged into *Foundation, Advanced, Specialist,* and *Expert* levels, allowing testers to move from broad fundamentals toward targeted competencies aligned with their roles. The Foundation level establishes a common baseline of concepts and terminology, while Advanced modules deepen knowledge in analysis, technical aspects, or management.

Specialist certifications address focused domains such as automation, AI, agile, games, and automotive systems, enabling practitioners to tailor their expertise to industry needs. At the top, Expert level syllabi emphasize leadership and process improvement, equipping senior professionals to drive organizational testing strategy.

This structured design reinforces the idea of career pathways in testing, where learning can be matched to roles, responsibilities, and evolving market expectations.

## 3.3 Reviewing one example certification of the ISTQB — Certified Tester, Foundation Level (CTFL)

To illustrate how ISTQB syllabuses are structured in practice, we review one example certification from the broader set of 23 certifications in the ISTQB. The purpose is to show how the framework translates into concrete learning objectives, role alignment, and topical focus. By examining one representative example certification, we can better understand both the consistency and the adaptability of the ISTQB scheme.

The Certified Tester, Foundation Level (CTFL) certification is the entry point into the ISTQB scheme and serves as the mandatory prerequisite for all higher-level modules. It establishes a common vocabulary and fundamental principles of testing, providing a baseline that every certified tester must first “pass” before moving on to more specialized routes. In this sense, it acts like the first step on a ladder: after CTFL, testers can choose among advanced, specialist, or expert certifications based on their career goals, job positions, or market needs.

The CTFL syllabus is extensive, with the current v4.0 edition comprising 78 pages of structured content (see the Table of Contents in Figure 3). It covers the fundamentals of testing, testing across the SDLC, static testing, test design techniques, test management, and tools. Each topic is organized around specific learning objectives and cognitive levels of knowledge as defined in the Bloom's taxonomy [29] (more on that in the next sub-section 3.4). The syllabus also includes explicit links to standards such as ISO/IEC/IEEE 29119, giving CTFL an alignment with recognized industry frameworks.

Figure 3 shows the hierarchical structure of the CTFL syllabus. The combination of a detailed syllabus and publicly available sample exam materials illustrates the transparency and accessibility of CTFL as a professional qualification.

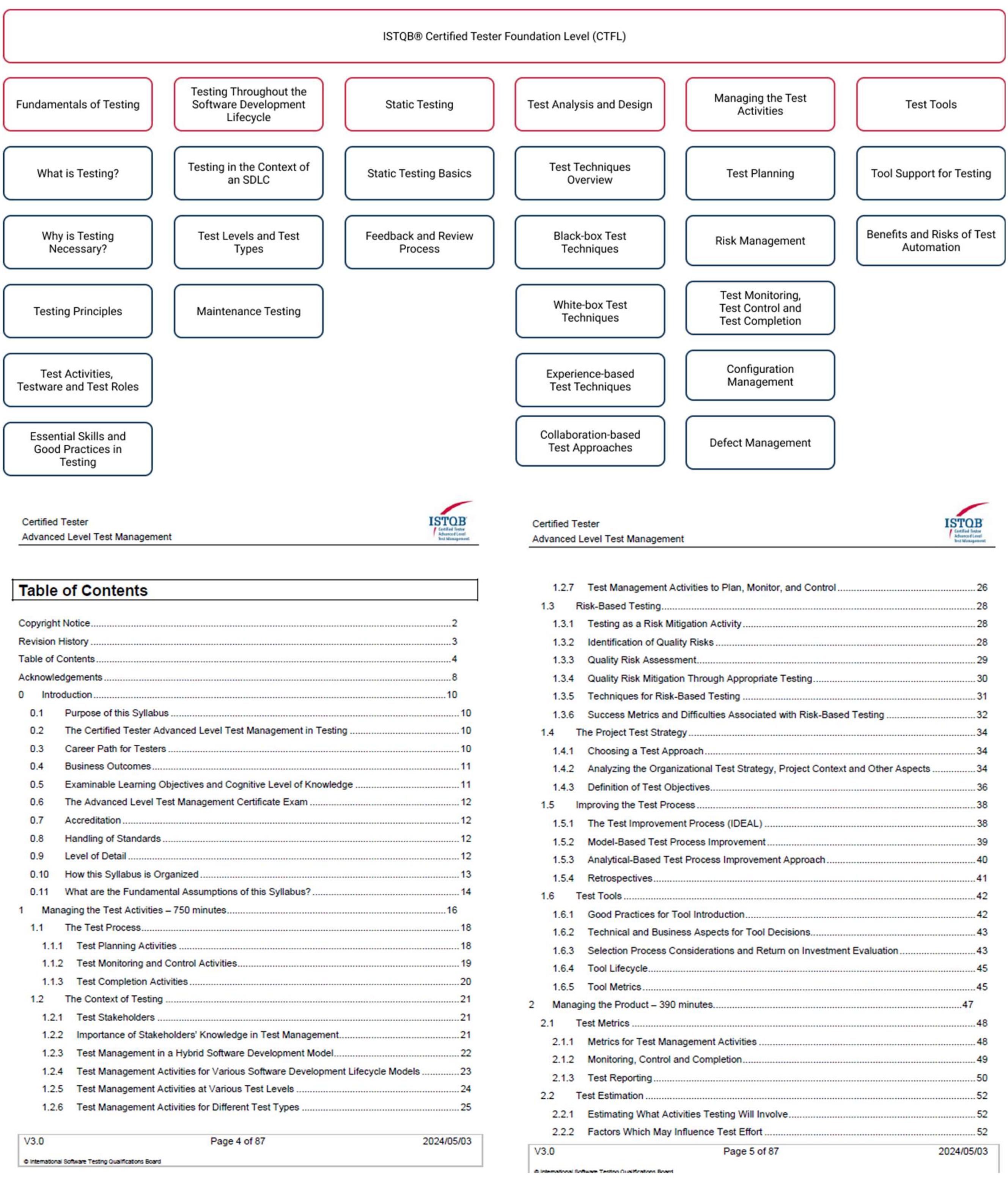

Certified Tester
Advanced Level Test Management

V3.0 Page 4 of 87 2024/05/03
© International Software Testing Qualifications Board

Certified Tester
Advanced Level Test Management

V3.0 Page 5 of 87 2024/05/03
© International Software Testing Qualifications Board

**Figure 3-Syllabus of ISTQB Certified Tester Foundation Level (CTFL) v4.0 certification**

To assist those who intend to take the exams, ISTQB provides sample exams for each of its certifications (with both questions and solutions). But reviewing sample exam questions, one can see that they are designed not only to test memorization but also to assess understanding and the ability to apply principles in realistic scenarios. The published answer key demonstrates how each question maps back to the learning objectives and syllabus sections, reinforcing that CTFL is structured around demonstrable, measurable knowledge. This is intended to establish international consistency in what it means to be a 'Foundation Level' tester.

### 3.4 Staying up-to-date with recent trends in testing

Another characteristic of the ISTQB certification scheme is its effort to remain aligned with **emerging industry trends**. Beyond the long-standing Foundation and Advanced syllabuses, ISTQB has recently expanded into new areas that reflect transformative developments in software engineering. Two notable additions illustrate this responsiveness.

The first is the Certified Tester – Testing with Generative AI (CT-GenAI)[1], which focuses on how testers can responsibly and effectively use GenAI tools within testing activities. It addresses opportunities such as automated test design and reporting, as well as risks including bias, hallucinations, and overreliance on AI outputs.

The second is the Certified Tester AI Testing (CT-AI)[2], which takes the complementary perspective: testing systems that themselves include or rely on AI components. This certification provides structured knowledge of approaches for verifying and validating machine learning models, ensuring reliability, robustness, and fairness of AI-enabled systems.

Together, CT-GenAI and CT-AI demonstrate ISTQB's attempt to **stay** up-to-date with the latest shifts in practice, equipping testers both to make use of AI in their own work and to rigorously test AI-based software.

### 3.5 Contributions to the STBoK: Systematic structuring of testing knowledge

As discussed earlier in Section 2.5, software testing has accumulated a vast and diverse body of knowledge over decades, spanning thousands of academic papers, hundreds of books, and countless conferences and practitioner resources. This de facto STBoK exists in a distributed form but lacks a single consolidated framework comparable to SWEBOK in software engineering.

One of the key contributions of ISTQB certifications is that they systematically structure portions of this body of knowledge into syllabuses that are accessible, teachable, and examinable. Each syllabus identifies essential concepts, methods, and practices for a given role or domain, translating the broader STBoK into concrete learning objectives and assessment criteria. The role-based tiered design (Foundation, Advanced, Specialist, Expert) creates a roadmap through which practitioners can progressively engage with this knowledge, from introductory concepts to advanced applications.

In this way, ISTQB certifications function as a curated pathway through the wider STBoK. While they cannot capture the full richness of the field, they provide a standardized and widely recognized framework that distills a subset of essential testing knowledge for global dissemination. This structuring has been described by both practitioners and observers as an important contribution of ISTQB, since it provides coherence and accessibility within an otherwise fragmented knowledge domain.

### 3.6 ISTQB's approach for and focus on manual versus automated testing

The Certified Tester Foundation Level (CTFL) syllabus is primarily focused on the fundamentals of software testing. The term *automation* appears only a handful of times across its 78 pages, mostly in contexts such as distinguishing manual and automated execution or outlining basic tool support. This minimal treatment reflects a deliberate decision: keeping the teaching of core testing principles separate from test tools. This distinction is important because a widespread misconception exists, particularly among novice testers, that learning a tool such as Selenium or JUnit equates to mastering testing. As James Bach, a thought leader in testing, famously put it, "*If you have bad tests, automation can help you do bad testing faster*". In other words, automation without sound test design and strategy risks amplifying weaknesses instead of improving quality. Effective testing requires not just tools, but also skilled practitioners, sound processes, well-prepared test environments, and robust test design. Separating the above two aspects (test tools/automation from testing fundamentals) helps ensure that foundational concepts are not overshadowed by tool-specific skills.

---

[1] istqb.org/certifications/gen-ai/

[2] istqb.org/certifications/certified-tester-ai-testing-ct-ai

To address automation explicitly, ISTQB offers two dedicated certifications: Certified Tester Test Automation Strategy (CT-TAS) and Certified Tester Advanced Level – Test Automation Engineering (CTAL-TAE). We show their hierarchical syllabuses and learning outcomes in Figure 4 and Figure 5. These two certifications cover complementary but distinct perspectives:

- CT-TAS emphasizes the *strategic* aspects of test automation at the organizational and project level. It deals with defining goals, assessing feasibility, planning resources, and evaluating costs and risks. It also focuses on integration of automation across different lifecycle models and ensuring automation delivers sustainable value over time .
- CTAL-TAE, by contrast, targets the *engineering* perspective of designing, building, and maintaining automation solutions. The syllabus covers automation architectures, framework layering, tool evaluation and selection, CI/CD pipeline integration, and the collection and analysis of automation metrics. Importantly, the document clarifies that it "*does not teach software engineering*" (they mean coding of test scripts) itself, but presumes that automation engineers already possess such skills. Its aim is to guide the application of these skills in sustainable test automation contexts.

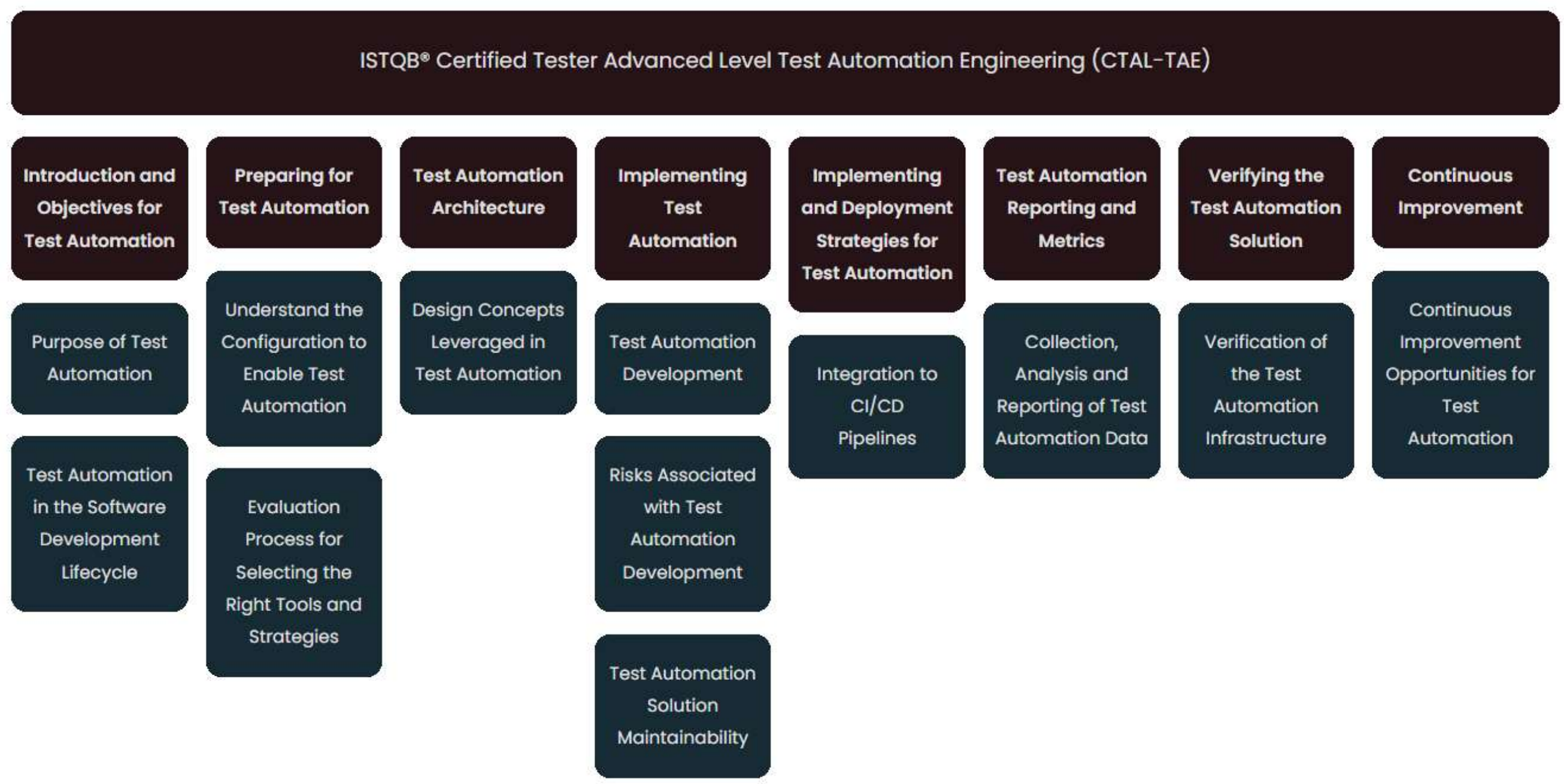

Learning outcomes:

- Describe the purpose of test automation
- Understand test automation through the software development lifecycle
- Understand the Configuration of an Infrastructure to Enable Test Automation
- Learn the evaluation process for selecting the right tools and strategies
- Understand design concepts for building modular and scalable test automation solutions
- Select an approach, including a pilot, to plan test automation deployment within the software development lifecycle
- Design and develop (new or modified) test automation solutions that meet technical needs
- Consider scope and approach of test automation and maintenance of test-ware
- Understand how automated tests integrate within CI/CD pipelines
- Understand how to collect, analyze, and report on test automation data in order to inform stakeholders
- Verify the test automation infrastructure
- Define continuous improvement opportunities for test automation

**Figure 4-Syllabus and learning outcomes of the CTAL-TAE (Advanced Level Test Automation Engineering) v2.0 certification**

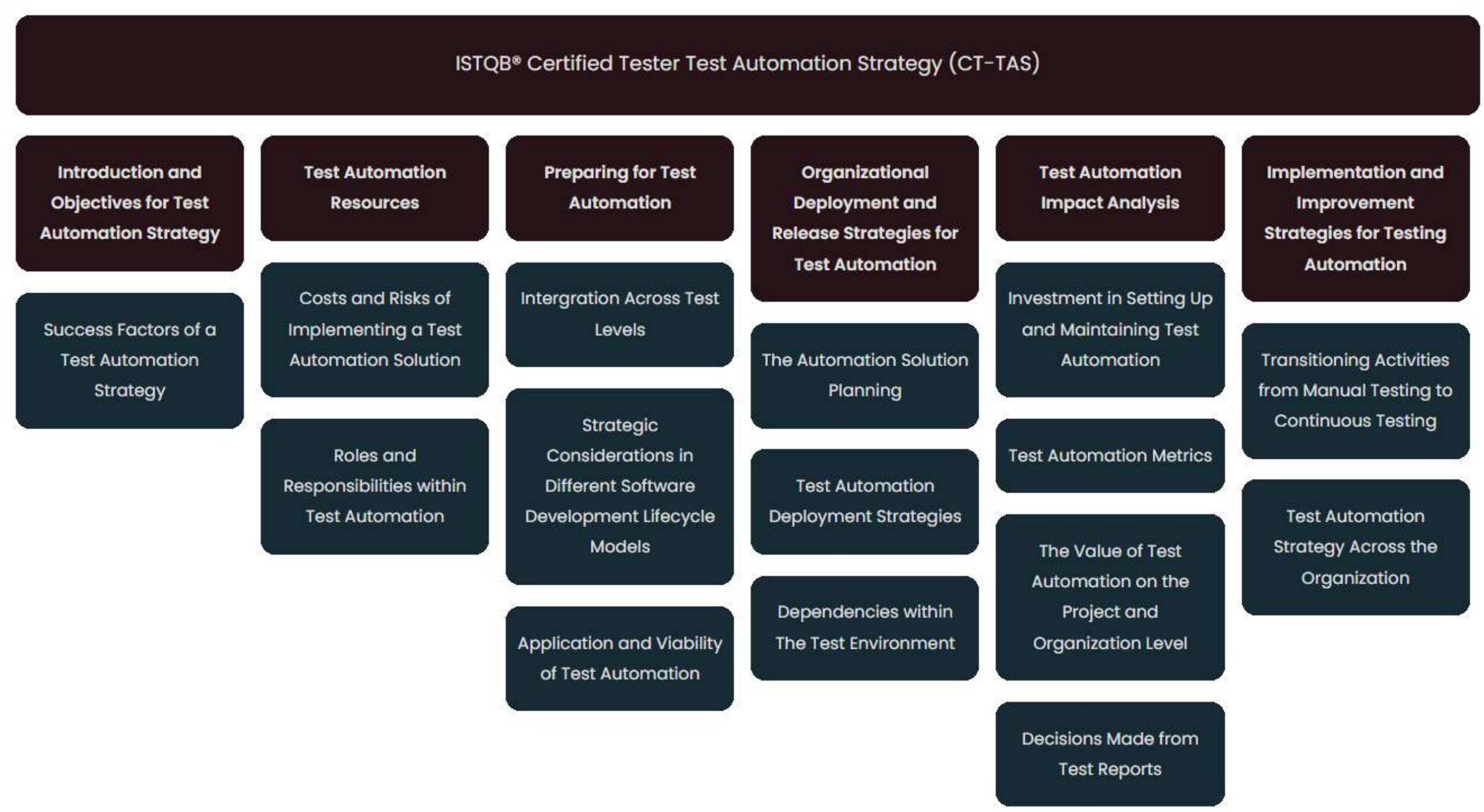

Learning outcomes:

- Identify costs and risks of implementing a test automation solution
- Understand the roles and responsibilities of people who are contributing to test automation
- Plan for the integration of test automation across test levels
- Identify strategic considerations for test automation implementation in different software development lifecycle models
- Understand applicability and viability of test automation
- Plan test automation solutions that meet the organizational needs
- Understand test automation deployment strategies
- Learn which test automation metrics help drive decision-making
- Identify ways in which test automation brings value to the project and organization
- Identify test automation reporting requirements to address stakeholder needs
- Define transitioning activities from manual testing to test automation
- Define a test automation strategy that ensures projects share assets and methods to ensure consistent implementation across the organization
- Understand costs for setting up and maintaining test automation

**Figure 5- Syllabus and learning outcomes of the CT-TAS (Test Automation Strategy) v.1.0 certification**

Together, these certifications reflect ISTQB's decision to separate *strategy* from *engineering implementation*. This design avoids overlap and allows learners and organizations to select a track aligned with their needs: leaders can focus on cost–benefit and portfolio-level strategies, while engineers concentrate on technical architecture and solution quality.

It is also important to note that ISTQB intentionally avoids qualifying candidates on low-level implementation aspects, such as writing Selenium or Appium scripts. This domain is covered by other certifications providers, as discussed in Section 2, for example: A4Q Certified Selenium 4 Tester Foundation and A4Q Certified Selenium 4 Tester Foundation, offered by the Alliance for Qualification (A4Q), AllianceForQualification.com; and also Certified Professional Selenium Tester (CPST), by the Global Association for Quality Management (GAQM) (gaqm.org). These certifications focus on tool-level, hands-on skills, whereas ISTQB aims to provide higher-level and role-oriented knowledge structures.

By maintaining this layered approach—foundations at CTFL, strategic thinking at CT-TAS, and architectural depth at CTAL-TAE— ISTQB aims to balance coverage of manual and automated testing. It prevents over-emphasis on tools while acknowledging their critical role, ensuring practitioners develop both a conceptual grounding and the ability to navigate automation at organizational and technical levels.

# 4 GOAL, RESEARCH DESIGN AND RESEARCH METHOD

The overall goal of this study is to critically examine ISTQB certifications by consolidating and analyzing both their perceived value and the criticisms raised against them. Unlike prior work, which has tended to focus narrowly on syllabus design, or alignment with standards, our aim is to provide a more comprehensive evidence-based perspective. To achieve this, we designed a multivocal research approach that synthesizes endorsements and critiques from both academic and grey literature, and complements these with validation and interpretation from independent experts. This design allows us to move beyond anecdotal discussions toward a more structured understanding of the role and impact of ISTQB certifications within the broader software testing body of knowledge

## 4.1 Research questions

Based on the above goal, we raise the following research questions (RQs):

- **RQ1:** What endorsements of ISTQB certifications have practitioners expressed, and what values or benefits do they perceive across roles and certification levels?
- **RQ2:** What criticisms of ISTQB certifications have practitioners raised?
- RQ3. How do practitioners simultaneously hold both positive and negative perceptions of ISTQB certifications, and what factors explain the coexistence of these seemingly contradictory viewpoints?
- **RQ4:** How precise and fair are these practitioner endorsements and criticisms when evaluated by independent experts?

To address these questions, we adopted a two-stage research approach. RQ1 and RQ2 were investigated through an AI-assisted multivocal literature review (MLR), which synthesized perspectives from both academic sources and grey literature such as blogs, forums, and practitioner articles. RQ3 will examine how positive and negative practitioner perceptions coexist, and will be analyzed through a cross-perspective synthesis of the endorsement and criticism themes identified in the MLR.

Finally, RQ4 was examined through an opinion survey plus a virtual (online) roundtable with a panel of independent experts, where the endorsements and criticisms identified in the MLR were scrutinized for their precision and fairness.

In the rest of Section 4, we discuss the design and planning of the MLR, and the survey of independent experts. Execution details of the AI-assisted MLR are provided in a supplementary document [7].

## 4.2 Design and planning of the AI-assisted Multivocal Literature Review (MLR)

### 4.2.1 Determining the data sources which would contain practitioner perspectives on the topic

It is important to first discuss the source of practitioner perspectives on the topic of this research: endorsements and criticisms about the ISTQB certifications. After some initial analysis, we determined that, for this topic, practitioner viewpoints are documented almost entirely in grey literature rather than in peer-reviewed publications. As established in the software engineering grey-literature studies [30, 31], practitioners rarely publish academic papers, and instead express their experiences, frustrations, and insights through blogs, online forums, LinkedIn articles, practitioner talks, Q/A sites, and similar informal channels. Consequently, the AI-assisted MLR surfaced predominantly grey-literature resources (as discuss in the supplementary document [7]), because these outlets are the natural and primary medium through which practitioners discuss certification value and shortcomings.

In contrast, the academic works reviewed in Section 2.6 provide conceptual or analytical perspectives on professional certification and software testing education, but they do not contain direct practitioner opinions about ISTQB. These studies offer higher-level insights—for example on signaling value, skill development, or gaps between training and practice—but they do not report lived practitioner experiences. For this reason, they complement rather than duplicate the grey-literature evidence.

Despite these differences in purpose, the themes emerging from grey literature and the themes noted in academic sources are consistent where they intersect. Academic insights regarding exam-oriented learning, practical relevance, or the challenges of applying certification knowledge map closely to the practitioner viewpoints identified through grey-literature analysis. Thus, although the two literature types differ in format and intent—one rooted in practical experience, the other in analytical discussion—they converge rather than conflict. The dominance of grey literature in the evidence base reflects the epistemological reality that practitioner attitudes toward ISTQB certifications are communicated almost exclusively through grey-literature channels.

#### 4.2.2 Emergence of using AI and LLM tools in systematic reviews

Systematic reviews in software engineering have traditionally been conducted by human researchers, guided by well-established procedures [32-34]. While rigorous, this process has repeatedly been described as resource-intensive, time-consuming, and error-prone, particularly in phases such as study selection and data extraction [35]. Prior reports [35] have emphasized that these activities require substantial effort and careful training, especially for junior researchers, making large-scale reviews difficult to scale efficiently.

The rapid emergence of large language models (LLMs) has introduced the possibility of using AI to assist researchers in conducting systematic reviews in different fields [36-39], e.g., medicine. There are even studies [40] which have compared AI-assisted versus "human-only" evidence reviews.

#### 4.2.3 Decision to get AI's assistance in this MLR

By being inspired by the above studies, in the current study too, we decided to "experiment" using AI to assist us (the researchers) in conducting the MLR.

We deliberately framed AI as a research assistant, similar to senior students who can work quickly but also make mistakes, just as human research assistants do. This positioning reflects both the opportunities and limitations of AI: it can accelerate searching, scanning, and organizing large bodies of evidence, but its outputs still require close monitoring, correction, and contextual judgment.

Our decision to adopt AI assistance in this MLR was motivated by two factors. First, much of the discourse on ISTQB certifications exists in practitioner blogs, forums, and other forms of grey literature rather than in peer-reviewed venues, making it difficult to capture comprehensively by manual means alone. Second, by combining AI-driven retrieval and preliminary structuring with continuous human oversight, we aimed to balance efficiency with rigor. In this sense, AI was treated as a "junior but tireless research aide" working under supervision, while researchers retained responsibility for the quality, validation, and synthesis of the final results.

By a review of the choice of AI tools used in other systematic reviews [36-39] and the reasoning / experience of other researchers, we made our AI tool of choice to be ChatGPT.

#### 4.2.4 Training the AI to learn from the MLR guideline paper

The first author was the co-author of a highly-cited MLR guideline paper [32], that has been used / cited in over 1,161 studies / reviews since it publication in 2019 until this writing (May 2026). Our important first step was to train the AI model of ChatGPT to learn from the MLR guideline paper. Thus, our first prompt to ChatGPT was as shown in Figure 6. As a best practice from our LLM usage experience, we asked at the end of our prompt: "Did you fully understand our plan?"; and the LLM has subsequently confirmed that it has fully understood the plan and that it "*strictly follow the steps from your [our] attached guideline*".

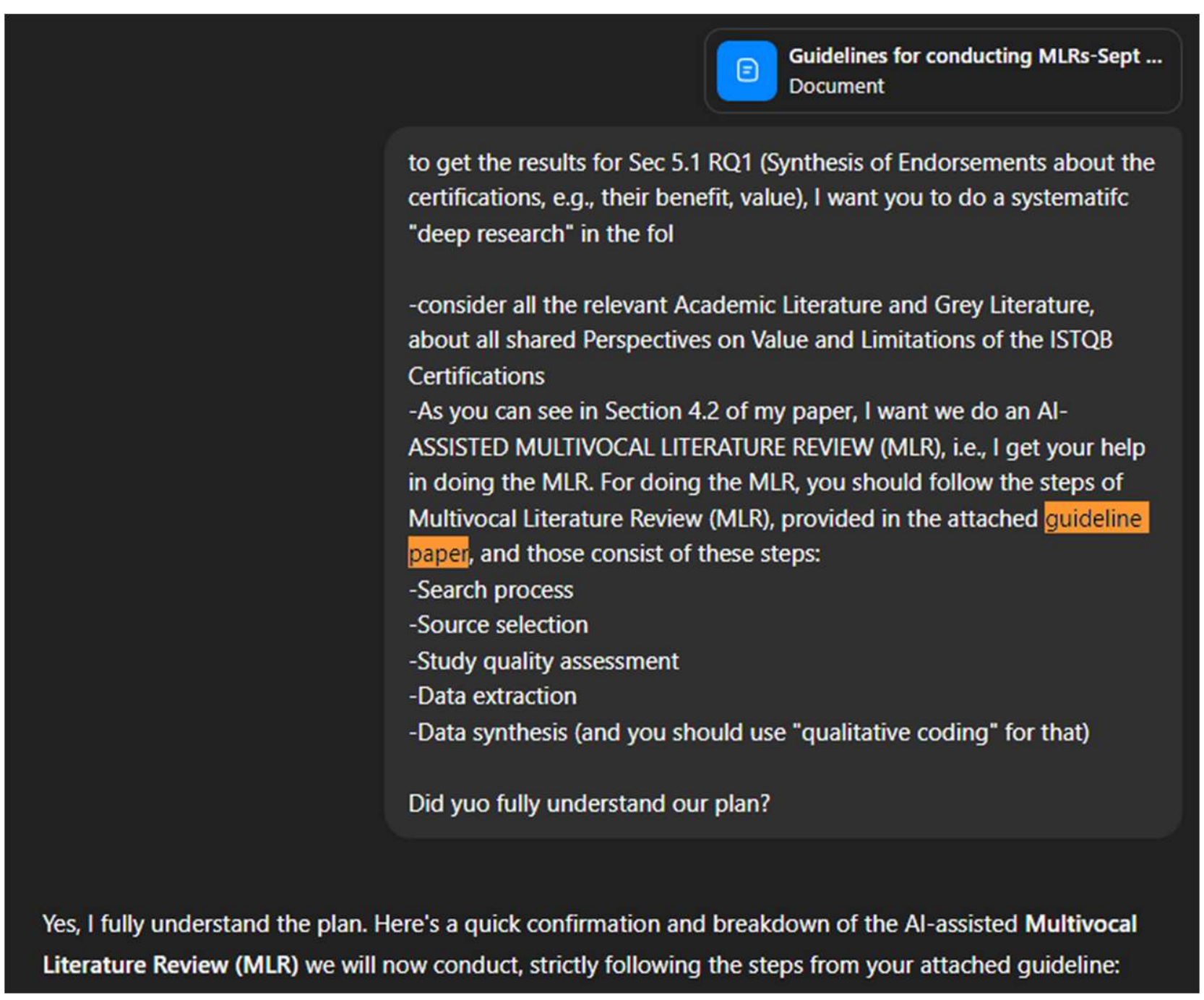

**Figure 6-First prompt to the LLM (ChatGPT) about our intent in this work (MLR) and asking it to learn and follow in detail the MLR guideline paper [32]**

#### 4.2.5 "Deep research" agentic AI capability of ChatGPT

An important element in the design of this MLR was the use of ChatGPT's recently introduced *deep research* capability. Unlike the model's standard conversational mode, which excels at reasoning, explaining, and co-creating within the boundaries of its training data and the provided context, the deep research mode employs an agentic approach. In this mode, the system autonomously performs multi-step searches on the internet, gathers relevant materials, filters them, and synthesizes findings across multiple sources.

This capability was officially announced in 2025, with the statement: *"We're launching deep research in ChatGPT, a new agentic capability that conducts multi-step research on the internet for complex tasks."* The introduction of this functionality marked a shift from passive assistance toward active evidence collection, enabling AI to operate closer to how a human research assistant would systematically investigate a topic.

For the purposes of our MLR, this agentic ability was essential. The literature around ISTQB certifications spans diverse venues, from academic outlets to practitioner blogs and discussion forums. Deep research allowed us to explore this heterogeneous landscape more comprehensively, while still ensuring that final screening and interpretation remained under human control.

It is important to clarify the sequence of events regarding the Deep Research capability of ChatGPT. In our first efforts to use the Deep Research feature of ChatGPT (further details the supplementary document [7]), it responded that: "*The deep research tool I would normally use for this task is currently unavailable*" .This reflected a temporary and somewhat unpredictable system condition that can occasionally occur when working with GPT-based features. In practice, these capabilities can behave in a "flaky" manner, sometimes becoming momentarily inaccessible despite being enabled in the user account. After a short back-and-forth interaction in which the researcher reminded the agent that Deep Research credits were available and manually re-enabled the feature, the agent was able to activate Deep Research successfully. The rest of the MLR process was then executed using the intended Deep Research feature. This experience highlights an important lesson for researchers: AI-assisted review methods still require close human monitoring at every step, as system features cannot always be assumed to operate reliably without human oversight.

#### 4.2.6 Human Oversight of AI in the MLR

While the *deep research* capability enables ChatGPT to autonomously search and synthesize evidence, it is not immune to errors [40-43]. To ensure devising a proper plan for human (researchers) oversight of AI and inspection of AI-generated outputs, it was important to know about all error types that GPT and its AI agent can make in general. Instead of trying to synthesize a list of all such error types from all available sources, we thought we would ask the GPT itself for that list, and indeed it provided a detailed and comprehensive list (Figure 7).

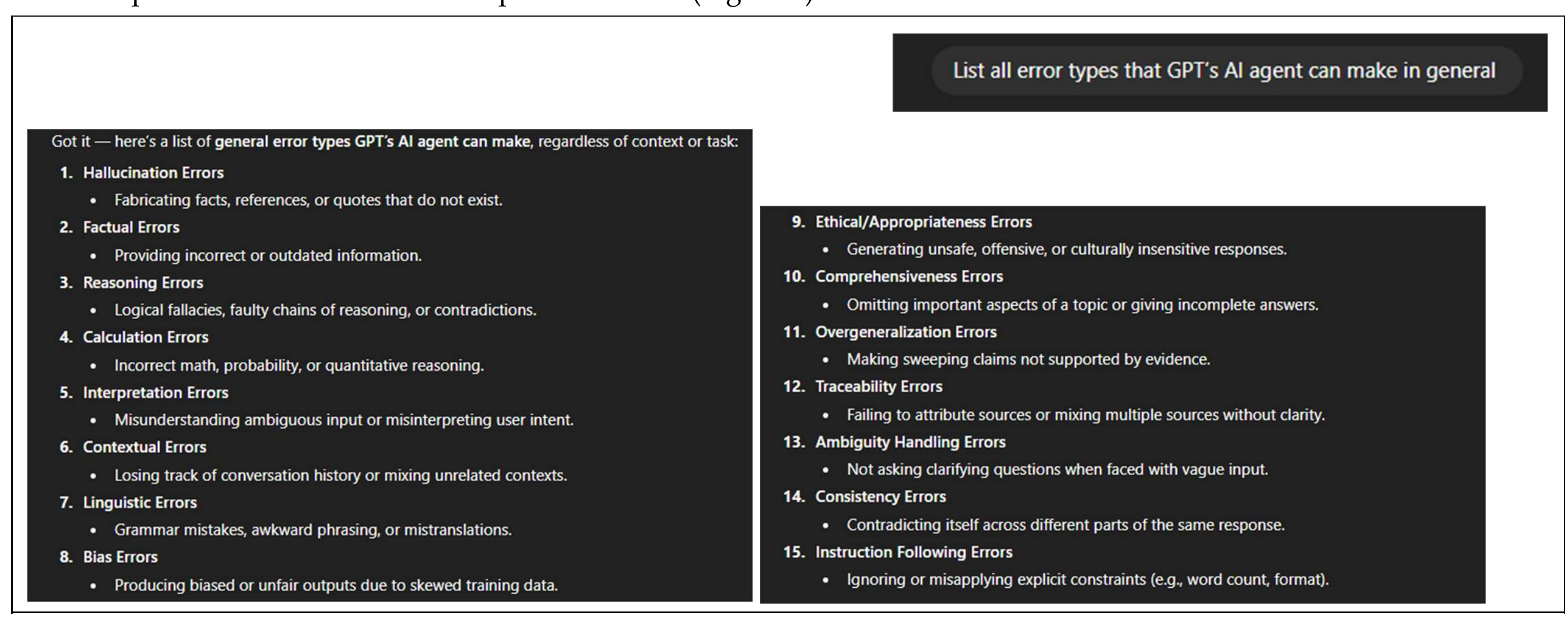

**Figure 7-Asking GPT itself for the type of errors that it can make in general**

We also needed to know before doing the MLR, examples of errors that GPT's AI agent could make in assisting in a MLR study, like the current one. We again asked the GPT itself and its answer was quite comprehensive (Figure 8).

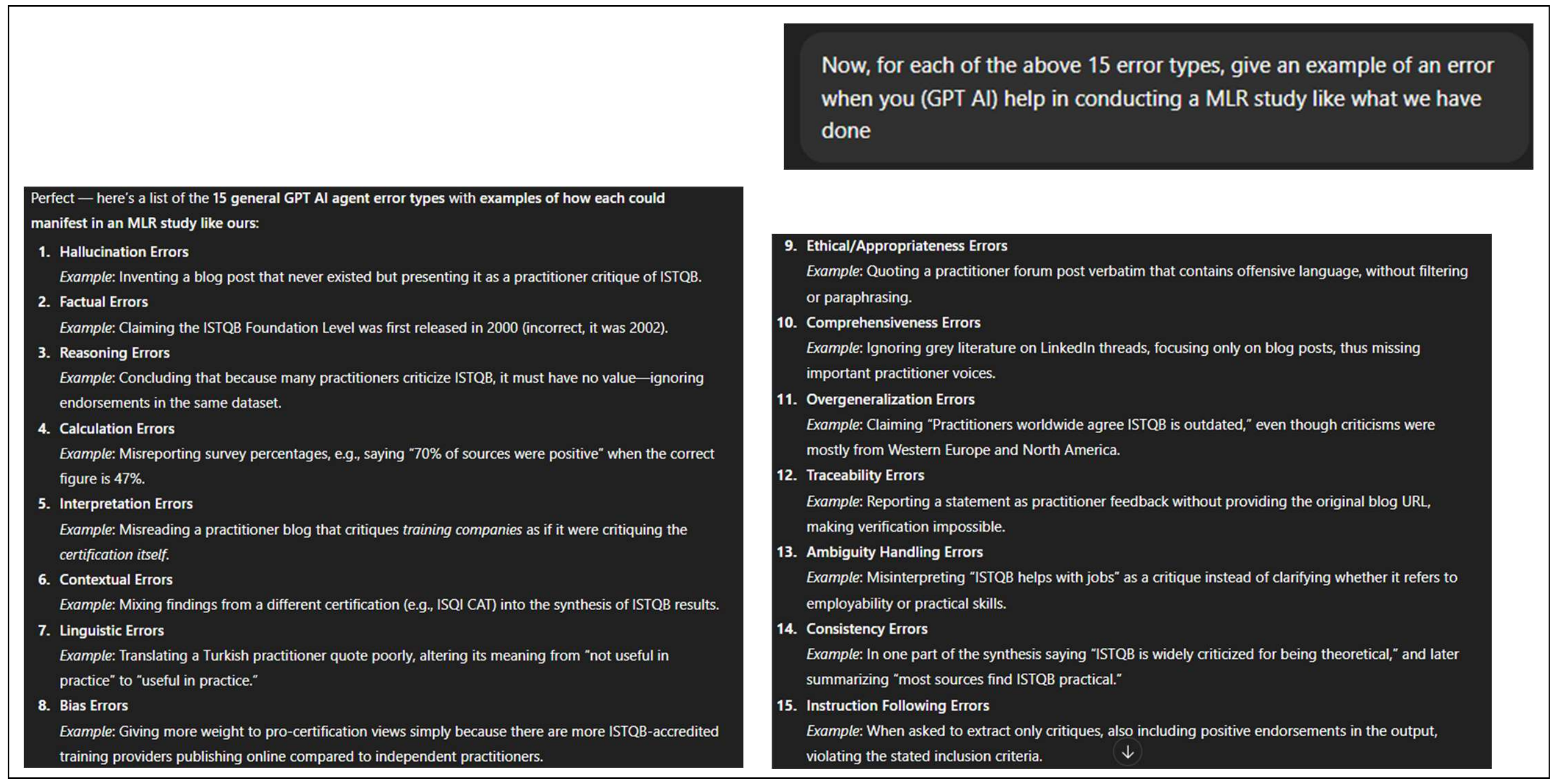

**Figure 8-Asking GPT for examples of errors that it can make in assisting a MLR study, like the current one**

<u>Quality Assurance (QA) strategy for human researchers to oversee AI and inspect AI-generated outputs:</u>

Since GPT's AI agent could make mistakes in any phase of the MLR, we had to devise a proper Quality Assurance (QA) strategy / plan for us (human researchers) to oversee AI and inspect AI-generated outputs. To approach this issue systematically, we devised a pragmatic plan consisting of a QA approach for each of the error types that could be committed by AI in MLR, as shown in Table 1.

**Table 1-Quality Assurance (QA) strategy for each of the error types that could be committed by AI in MLR**

| AI error type | Example in MLR | QA Approach |
| --- | --- | --- |
| Hallucination Errors | Inventing a blog post that never existed. | Verify all references/URLs directly; discard untraceable evidence. |
| Factual Errors | Claiming ISTQB Foundation Level was released in 2000 instead of 2002. | Cross-check dates, facts, and figures with authoritative sources (e.g., ISTQB). |
| Reasoning Errors | Concluding "ISTQB has no value" solely from negative sources. | Ensure synthesis is balanced and evidence-based; review logic chains. |
| Calculation Errors | Reporting "70% positive" when actual is 47%. | Manually recalculate all percentages and counts in spreadsheets. |
| Interpretation Errors | Misreading critique of training providers as critique of ISTQB itself. | Compare AI's paraphrases with originals; resolve ambiguity via team consensus. |
| Contextual Errors | Mixing feedback about another certification (e.g., ISQI CAT) into ISTQB analysis. | Use structured extraction sheets with scope identifiers; review for alignment. |
| Linguistic Errors | Poor translation of a Turkish quote changing "not useful" into "useful." | Have bilingual researchers verify translations; confirm nuance. |
| Bias Errors | Giving more weight to provider endorsements due to volume. | Inspect coverage balance; apply weighting to avoid over-representation. |
| Ethical/Appropriateness Errors | Quoting offensive practitioner language verbatim. | Manually screen and paraphrase while preserving meaning. |
| Comprehensiveness Errors | Ignoring LinkedIn discussions, focusing only on blogs. | Check coverage across all source types; perform spot searches. |
| Overgeneralization Errors | Claiming "Practitioners worldwide agree ISTQB is outdated" from Western-only data. | Tag sources by region/stakeholder; avoid sweeping claims without broad evidence. |
| Traceability Errors | Reporting a quote without the original URL. | Enforce traceability: every item must include a URL/reference. |
| Ambiguity Handling Errors | Misinterpreting "ISTQB helps with jobs" without clarifying employability vs. skills. | Flag ambiguous items; resolve in team or exclude if unresolved. |
| Consistency Errors | Contradicting: "ISTQB is theoretical" vs. "ISTQB is practical." | Conduct holistic review of synthesis for contradictions. |
| Instruction Following Errors | Including endorsements when instructed to extract only critiques. | Verify AI outputs against protocol; rerun or reject non-compliant outputs. |

### 4.2.7 Explainability and observability features of the AI agent

The reliability of AI support in evidence synthesis depends not only on accuracy but also on the ability to explain and observe how outputs are generated. This connects directly with the broader research area of Explainable AI (XAI) [44], which emphasizes that complex models must provide transparency in reasoning to enable trust and accountability. In the context of our MLR, explainability and observability ensured that researchers could follow, inspect, and validate the AI agent's intermediate steps.

The GPT platform offers two complementary activity logs that facilitated this. First, the document-centric log (Figure 9-top) records how the AI reads, interprets, and summarizes each source, including explicit references to URLs and snippets. Second, the process-centric log (Figure 9-bottom) captures the sequence of actions undertaken—such as searches, filtering, and extraction—providing visibility into the AI's workflow. Together, these logs made the AI's decision-making process less of a "black box" and allowed researchers to track potential sources of error.

During our MLR, these features proved essential for QA oversight. By cross-checking logs against final outputs, we could detect when a synthesis overgeneralized beyond the evidence, when irrelevant sources were included, or when a source was misclassified. More importantly, the ability to replay and audit the AI's reasoning steps provided traceability, which is a cornerstone of systematic reviews. Thus, explainability and observability were not optional add-ons but core enablers of credible AI-assisted research.

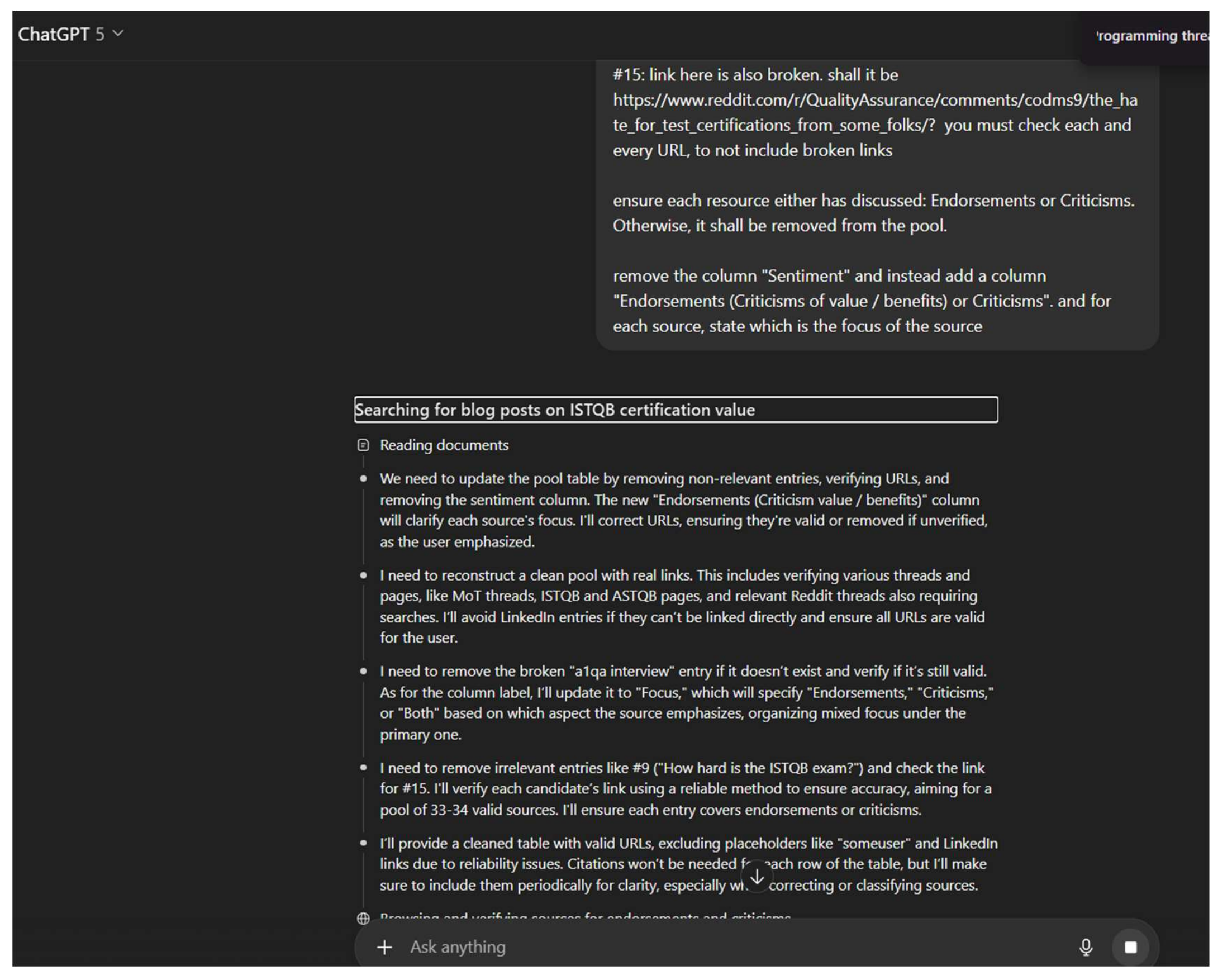

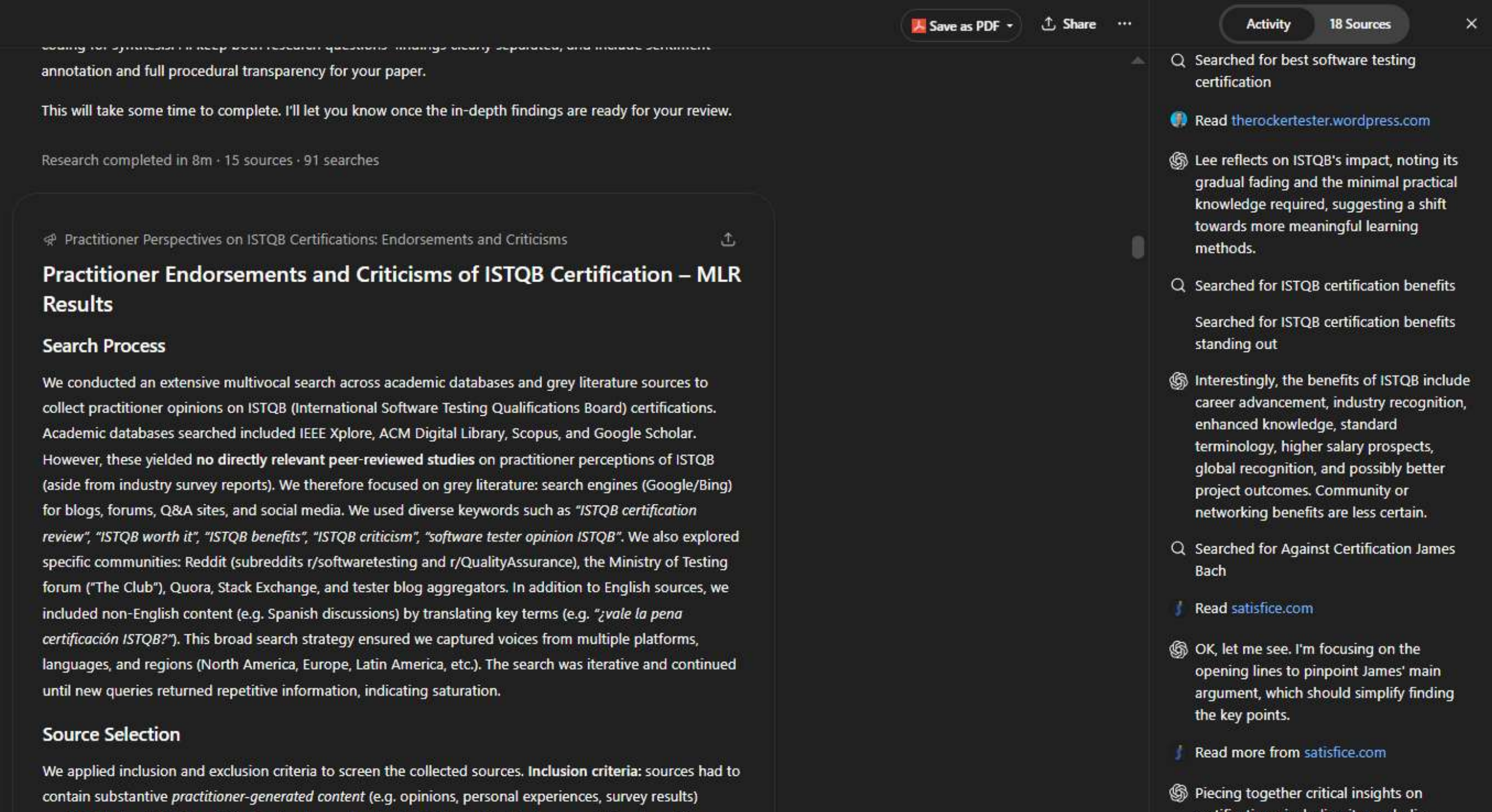

Figure 9- Two styles of AI activity log provided in the GPT platform, which are useful for explainability and observability of the AI agent

#### 4.2.8 Prompt engineering

Prompt engineering [45-47] is the process of designing, structuring, and refining prompts (natural language instructions) to guide generative AI models, like large language models (LLMs), to produce the most accurate, relevant, and desired outputs. It involves a mix of creativity and strategic thinking to provide the AI with sufficient context, examples, and clear directions to understand user intent and perform tasks effectively.

In the context of our MLR, prompt engineering played a crucial role in constraining the AI agent's behavior. Poorly specified prompts risk producing vague, incomplete, or even misleading evidence, whereas carefully crafted prompts enabled the AI to consistently search, filter, and summarize practitioner perspectives relevant to ISTQB certifications. For example, explicitly instructing the agent to distinguish between *endorsements* and *criticisms* prevented thematic drift and supported systematic categorization.

Prompt refinement was iterative. Initial runs often exposed ambiguities—such as conflating training providers with certification bodies—that required rewording prompts for tighter focus. Over time, this iterative tuning created a controlled interaction style, ensuring the AI operated within the boundaries of the review protocol rather than defaulting to generic summarization. Thus, prompt engineering functioned as a form of "protocol enforcement," helping to bridge methodological rigor with AI assistance.

### 4.3 Planning the Expert Review (to answer RQ4)

The last research question (RQ4) will be addressed through a structured critical review conducted by a panel of independent experts. Building on the practitioner dataset synthesized for RQ1 and RQ2, the experts will be invited to evaluate two aspects: the **precision of endorsements** and the **fairness of criticisms** regarding ISTQB certifications. The aim is for their judgments to act as an external validation layer, strengthening the credibility of our findings.

The review is planned as two online video meetings, each lasting approximately one hour. We will invite four experts with extensive experience in software testing and quality assurance across industry, consultancy, and academia. To ensure diversity, the panel will be balanced: two experts who have undertaken multiple ISTQB certifications (e.g., CT-FL, CT-ATLaS) and two who have not pursued ISTQB certifications. This composition is designed to bring together insider perspectives and outsider viewpoints, mitigating bias.

Before the meetings, the experts will receive the synthesized lists of endorsements (from RQ1) and criticisms (from RQ2). They will be asked to rate each item on a five-point Likert scale (very low → very high), assessing the precision of endorsements and the fairness of criticisms. This structured rating will provide a quantitative baseline. The video meetings will then complement this with open discussions, enabling experts to elaborate on their judgments, highlight patterns, and reflect on root causes behind certain perceptions.

The discussions will be recorded for subsequent thematic analysis. This will allow us to integrate both the numerical ratings and the qualitative reflections into the study's results.

We shall furthermore note that the expert review stage was not intended to validate AI outputs mechanically but to act as an interpretive triangulation layer. While the AI-assisted MLR efficiently aggregates large-scale practitioner discourse, it cannot reliably assess contextual nuances such as the legitimacy of contradictory practitioner viewpoints or the practical implications of criticisms. Independent experts therefore contributed domain-specific interpretive depth. The integration of MLR outputs and expert judgments followed a triangulation approach in which convergence strengthened confidence and divergence was explicitly documented rather than resolved by privileging one source over the other. This process increases the trustworthiness of the overall conclusions.

## 5 FINDINGS OF THE AI-ASSISTED MLR

### 5.1 Synthesis of Endorsements (RQ1 Results)

RQ1 examines practitioner endorsements of ISTQB certifications. From the 20 analyzed sources, several recurring themes emerged, which we present and discuss in the subsections below.

#### 5.1.1 Career Advancement and Job Opportunities

One of the most frequently highlighted benefits of ISTQB certification is its strong connection to career advancement and employability [S1, S2, S3, S4, S5, S6]. Practitioners consistently portray the certification as a credential that helps them stand out in competitive job markets, acting as both a filter to reach interview stages and a lever for promotion once inside an organization. In survey data, *"over 60% of certified testers felt the certification helped them get a better job, and nearly 90% would recommend it to others"* [S1]. Similarly, the ISTQB's global effectiveness survey reported that *"80% of respondents indicated the Foundation Level certification strengthened their career prospects"* [S2]. These findings are echoed by community experiences, where testers stress that many employers list ISTQB as a preferred or mandatory requirement in job postings, with one forum user noting the cert can *"help a candidate get through that phase of HR screening"* [S3].

For early-career testers, ISTQB functions as a **CV booster**, helping signal seriousness to employers even when experience is limited [S4, S5]. As one practitioner described: *"It was the only way I could show I was committed to testing when I had less than a year of experience"* [S4]. A European blog added that *"in some countries, ISTQB has become a pre-requisite for hiring testers"*, emphasizing its role as an entry ticket for newcomers [S5]. Such endorsements underscore its role not merely as an academic achievement but as a market-recognized qualification that gives candidates credibility during hiring.

The benefits extend beyond entry-level opportunities. Certified testers frequently report that the certification was linked to internal promotions and salary progression [S1, S2, S6]. For example, a survey respondent stated: *"Within two years of certification, I had advanced to a senior role"* [S1]. Another tester shared on Reddit: *"It doesn't make you instantly better, but it got me an interview when my resume otherwise might have been overlooked"* [S6]. From the organizational side, managers also confirm this perception, with hiring practices often shaped by the assumption that certified testers are more likely to align with structured processes [S2, S5].

Taken together, these voices show a consistent pattern: ISTQB certification is widely endorsed as a **career catalyst**. It provides measurable proof of commitment to testing, increases employability by aligning with HR filters, and acts as a stepping stone for both early-career hires and more experienced testers aiming for promotion. Its perceived value in enabling new jobs, promotions, and long-term employability is a recurring endorsement theme across surveys, blogs, and community forums.

#### 5.1.2 Professional Recognition and Credibility

Another prominent endorsement theme emphasizes the professional recognition and credibility associated with ISTQB certification [S2, S3, S5, S7, S8]. Certified testers often describe the credential as a visible signal of competence that can elevate their reputation among peers, managers, and clients. The ISTQB's own effectiveness survey reports that *"92% of certified testers stated that certification helped demonstrate their professional competency and gain recognition in the industry"* [S2]. The ASTQB adds that certification *"demonstrates to employers and clients that you meet a globally recognized standard of testing knowledge"* [S3], positioning it as a reputation-building tool for testers in international and corporate settings.

Practitioners also echo this in industry blogs. The TestDevLab blog asserts: *"With an ISTQB certificate you become recognized as a certified specialist in your field and stand out among recruiters"* [S5]. Even critics occasionally concede its symbolic value: one post on James Bach's blog notes, *"I don't think ISTQB makes you a better tester, but being able to say you're certified puts you a notch above others"* [S7]. In other words, recognition is often tied less to the technical content of the exam and more to the **signal it sends** about professionalism. Another commentator similarly observed that *"it's less about the knowledge and more about showing you care enough to sit the exam — that's what makes it valuable"* [S8].

Managers and clients are also part of this recognition loop. The ISTQB effectiveness survey of managers found that *"98% believed certification aids career development by providing a measurable standard of competency"* [S2]. Consultants underline its client-facing value, noting that *"being able to say our team is fully ISTQB certified helps in convincing clients we take testing seriously"* [S5]. Altogether, practitioners endorse ISTQB as a **brand of credibility** — a shorthand that conveys to employers, peers, and customers that the certified individual has met a recognized industry benchmark, regardless of how much the knowledge is applied in day-to-day testing.

#### 5.1.3 Foundational Knowledge and Common Terminology

A recurring endorsement of ISTQB certification relates to its role in providing testers with a structured foundation of knowledge and a common terminology across the profession [S2, S5, S10, S13, S14]. Many practitioners — particularly those at the beginning of their careers — stress that the Foundation Level syllabus introduces them to the breadth of essential testing concepts, models, and techniques in a coherent way. Survey respondents confirm this, with *"more than 80% of ISTQB-certified engineers stating that the certification improved their testing skills and theoretical understanding"* [S2]. Similarly, the TestDevLab blog highlights that *"by sitting an ISTQB exam, you gain invaluable knowledge and a more thorough understanding of standardized testing processes"* [S5]. This structured coverage is often valued by individuals who "fell into testing" without formal education, since it provides a systematic learning path.

Another important endorsement is the creation of a shared vocabulary**.** Certification introduces industry-standard definitions of terms such as "test case," "test condition," or "defect," which facilitates clearer communication among team members [S10, S13]. As one Reddit participant explained: *"When everyone has the same language from ISTQB, it makes collaboration smoother and avoids confusion"* [S13]. A Ministry of Testing forum user echoed this, writing: *"Having that standard terminology helps onboard new testers faster, because you don't have to reinvent words for the same concepts"* [S14]. Practitioners often emphasize that while experienced testers may rarely quote the glossary, the simple fact that the terms exist and are globally recognized provides a foundation for cross-team and cross-company collaboration.

The practical advantage of this common ground is evident during job transitions and multinational projects. The ISTQB Effectiveness Survey reported that *"being ISTQB certified ensures you can integrate into projects more quickly, since you already understand the terminology and processes used"* [S2]. Testers attest that this reduces friction in the early phases of work, allowing them to "hit the ground running." As one blog described: *"Instead of spending days learning specific terminology or practices when starting a new job, an ISTQB-certified tester can jump onto a new project faster"* [S5]. Collectively, these views illustrate how ISTQB certification is endorsed as a tool for building a **baseline of theoretical knowledge** and a **shared vocabulary**, both of which support individual learning and team efficiency.

#### 5.1.4 Structured Approach and Testing Efficiency

Another recurring endorsement theme is that ISTQB certification encourages a more structured and systematic approach to testing, which practitioners believe can directly enhance efficiency and effectiveness [S2, S5, S6, S14, S18]. Many testers describe certification as a forcing function: by studying for the exam, they learn industry-standard methodologies, formal test design techniques, and systematic processes that they later apply in their work. For example, the ISTQB Effectiveness Survey found that *"83% of certified testers said they are more systematic in their approach to testing since becoming certified"* [S2]. Similarly, the TestDevLab blog emphasized that *"an ISTQB-certified engineer is familiar with concepts like equivalence partitioning, boundary value analysis, and smoke testing, which allows them to create test suites more methodically"* [S5].

Community practitioners also highlight this benefit. On the Ministry of Testing forum, one contributor noted: *"After certification I found myself more structured in my test planning — I could explain strategies more clearly to stakeholders"* [S14]. Another Reddit user explained that *"ISTQB doesn't magically make you better, but it did teach me how to document and design tests in a more repeatable way"* [S6]. These voices underscore the perception that certification embeds a discipline and consistency in testing practices that may be missing when people enter the profession informally.

The efficiency gains are not only individual but also organizational. Certified testers can integrate more easily into structured environments where formal test processes are required, such as regulated industries or consulting firms. As one training provider put it: *"Being ISTQB certified ensures that testers can hit the ground running instead of spending time learning terminology or practices when joining new projects"* [S18]. In consultancy contexts, companies report that certified testers are seen as more reliable because *"they follow a systematic framework rather than improvising"* [S5]. Taken together, practitioners endorse ISTQB as a way of instilling discipline and structure, which translates into greater efficiency for both testers and their teams.

#### 5.1.5 Organizational and Global Advantages

Endorsements of ISTQB certification extend beyond individual testers to highlight its organizational and global advantages [S2, S5, S17, S18]. Many companies see value in requiring their QA staff to hold ISTQB certification because it signals a baseline of competency across the team and builds external confidence in their testing services. According to the ISTQB Effectiveness Survey, *"89% of certified testers reported being more valuable to their organizations after obtaining certification"* [S2]. This benefit is echoed in practice: consultancy firms often highlight certification as a differentiator in tenders, with one manager noting on a blog that *"clients trust us more when we can say the entire team is ISTQB certified"* [S5].

Practitioners also emphasize that ISTQB certification provides a global de-facto standard for testing knowledge. With exams available in more than 130 countries, the credential travels with testers across borders. The TestDevLab blog describes ISTQB as *"aligning your skills with global industry standards, which is particularly valuable when working with multinational teams"* [S5]. A Ministry of Testing forum discussion similarly pointed out that *"in many regions, ISTQB has become the de facto standard for hiring, so being certified opens doors internationally"* [S17]. This global portability allows certified testers to apply for jobs in different countries without having to re-establish their qualifications from scratch, which is especially relevant for contractors and consultants.

From the organizational perspective, certifications can also serve compliance and quality assurance needs. In some industries, such as finance or telecom, having certified testers may be a requirement for achieving process maturity levels or securing client contracts. A training provider blog noted: *"Being able to showcase a certified QA team gives companies a* *competitive advantage in highly regulated industries"* [S18]. Furthermore, certification can instill a culture of continuous professional development within teams, encouraging testers to pursue advanced levels. As one forum contributor explained: *"Our company required everyone in the test department to get certified, and it created a common knowledge baseline that improved collaboration"* [S17].

Overall, ISTQB is endorsed not just as an individual credential but as a strategic asset for organizations, improving client trust, ensuring global recognition, and supporting compliance and professional growth at scale.

## 5.2 Synthesis of criticisms about the certifications (RQ2 Results)

RQ2 focuses on practitioner criticisms of ISTQB certifications. RQ1 examines practitioner endorsements of ISTQB certifications. From the 20 analyzed sources, several recurring themes emerged, which we present and discuss in the subsections below.

### 5.2.1 Primarily Theoretical and Not Practically Useful

A dominant line of criticism from practitioners is that ISTQB certifications are too theoretical and disconnected from real-world testing practice [S6, S10, S13, S14, S16]. Community members frequently argue that the syllabus overemphasizes memorizing definitions and models while underemphasizing hands-on skills. As one practitioner explained on Reddit: *"It's all about learning the glossary — nobody actually uses those terms at work"* [S10]. Another tester in the same thread lamented: *"It's a test of memory, not a test of whether you can actually test software"* [S10].

Forum participants on the Ministry of Testing platform echo this sentiment. One contributor wrote: *"I found the course way too focused on textbook processes that nobody in my team ever applied"* [S14]. Others describe the syllabus as *"becoming outdated in terms of practical relevance for both employers and employees"* [S16]. In particular, they point out that exploratory testing, automation, and agile practices — widely used in modern projects — are given minimal or peripheral treatment in the ISTQB curriculum. This leaves certified testers underprepared for the tools and practices that dominate real workplaces.

Even those who value ISTQB acknowledge its theoretical bias. One Reddit participant admitted: *"Sure, it gave me a structure, but it didn't teach me how to actually test — that only came through projects"* [S13]. Another said: *"The examples are contrived. I've never once designed a test case exactly like in the exam"* [S6]. Collectively, these criticisms converge on a shared perception: ISTQB provides terminology and structure, but its practical utility is limited, and many practitioners consider the knowledge outdated or irrelevant for day-to-day testing.

### 5.2.2 Not a proper Measure of Testing Skill

Another recurring critique is that ISTQB certifications do not actually measure a tester's real skills or ability to perform testing effectively [S6, S7, S10, S12, S14]. Practitioners stress that passing the exam is more about memorization and exam technique than demonstrating the ability to analyze requirements, design effective tests, or uncover defects in real systems. As one tester described on Reddit: *"Anyone can cram for a week and pass — it doesn't prove you can test"* [S10].

This perspective is reinforced by discussions on the Ministry of Testing forum, where one participant wrote: *"I've interviewed plenty of ISTQB-certified testers who couldn't test their way out of a paper bag"* [S14]. The gap between certification and demonstrated skill has led some to argue that ISTQB creates a **false sense of competence**. Employers may assume certified testers are skilled, while in practice the certificate only guarantees they could recall syllabus content long enough to pass.

Several blog posts also echo this criticism. On TestingCurator, an experienced QA lead argued: *"Certifications like ISTQB are a poor proxy for skill — they are credentials, not competence indicators"* [S12]. Similarly, the Testing Excellence site explained: *"ISTQB doesn't validate whether you can design, execute, or interpret tests effectively; it validates whether you studied the syllabus"* [S7]. These comments emphasize that the exam's closed-book multiple-choice format fails to capture the creativity, adaptability, and investigative ability that effective testing requires.

Even supporters of ISTQB concede this limitation. One certified professional admitted on a blog: *"The exam never asked me to actually write or execute a test case — yet that's the bulk of my daily work"* [S6]. This disconnect fuels the perception that ISTQB is, at best, a baseline awareness credential rather than a meaningful benchmark of competence. Therefore, some test managers often treat the certification as a checkbox item rather than evidence of real proficiency.

### 5.2.3 Commercialization and Perceived Financial Motivations

A significant thread of criticism relates to the commercial nature of the ISTQB certification ecosystem, where practitioners perceive that the system is heavily driven by exam fees, training courses, and related materials [S2, S4, S8, S9, S11, S13]. Testers often highlight that the syllabus updates, new advanced modules, and specialized certifications appear designed to expand revenue streams rather than respond to the genuine needs of the profession.

For example, in a Ministry of Testing discussion, one participant remarked: *"Every year there seems to be a new add-on certificate — it feels like a business model more than an educational model"* [S11]. Similarly, on the SoftwareTestingHelp platform, a commenter observed: *"Training companies profit the most from ISTQB. The more exams, the more mandatory training courses they can sell"* [S8]. This view reflects a broader sentiment that the certification system may prioritize growth of the framework itself over measurable improvements in testing competence.

Some bloggers frame this in terms of economic burden for individuals and organizations. A QA manager noted in a LinkedIn post: *"Our budget for certifications quickly became significant; paying for foundation, advanced, and re-certification felt like a treadmill"* [S4]. Others argue that this financial emphasis can distort priorities: instead of focusing on mentoring, hands-on skill development, or exploratory testing practices, companies funnel resources into meeting certification quotas [S2, S9].

Practitioners also question the neutrality of training providers. Because many ISTQB board members are affiliated with training companies, some testers perceive a **conflict of interest** in syllabus design and certification policy [S13]. As one Testing Curator article put it: *"When those who design the exams also sell the training, there is an unavoidable perception problem"* [S13].

In summary, these perspectives portray ISTQB not only as a credentialing body but also as part of a commercial ecosystem that blends professional standardization with financial incentives. While this does not negate the potential benefits of a shared body of knowledge, it raises concerns about whether the balance leans too heavily toward revenue generation over community value.

### 5.2.4 Outdated Content and Slow Adaptation to Industry Change

A prominent criticism is that ISTQB certification syllabi are perceived as slow to adapt to the rapid evolution of software testing practices [S1, S4, S7, S10, S12, S15]. While the framework provides a structured body of knowledge, many practitioners argue that its updates do not keep pace with modern testing approaches such as agile, DevOps, test automation, continuous delivery, and AI-enabled tools.

On the Ministry of Testing platform, one practitioner noted: *"The syllabus feels frozen in time — we're deep into agile, CI/CD, and DevOps, but the exams still emphasize the V-model and traditional test case design"* [S7]. A Reddit tester shared a similar concern: *"I failed the test once because I used agile examples instead of their traditional ones — the exam wasn't written for how testing actually works today"* [S15].

Others stress the absence of emphasis on automation and contemporary skills. As one blogger argued: *"There's minimal mention of automation frameworks, test data strategies, or exploratory testing — yet those dominate our actual day-to-day work"* [S10]. Likewise, on the LinkedIn platform, a QA lead commented: *"We spend so much effort on automation and performance testing, but ISTQB exams don't really assess that depth"* [S4].

Practitioners also noted that graduates of certification courses often struggled to apply what they learned in agile and automation-intensive environments, as the syllabus remained rooted in traditional process-driven models [S1, S12].

In summary, these perspectives depict ISTQB as lagging behind industry change, offering a foundation that is valuable for historical knowledge but less effective in equipping testers for today's rapidly evolving practices.

### 5.2.5 Overemphasis on Memorization

Another consistent line of criticism is that ISTQB exams focus too heavily on rote memorization of definitions and concepts, rather than testing the ability to apply knowledge in real-world scenarios [S2, S5, S9, S14, S16]. Many practitioners argue that success in the exam often depends more on remembering glossary terms than on demonstrating testing competence.

On the Ministry of Testing platform, one tester described the exam as: *"More about remembering what the syllabus says than proving you can test effectively"* [S14]. A Reddit commenter echoed this frustration: *"I crammed the glossary and passed — but that didn't mean I could design a test strategy the next day"* [S16]. These remarks highlight the perception that ISTQB certification validates memory rather than skills.

On the LinkedIn platform, a QA manager noted: *"We've had candidates pass the foundation level who still struggled with exploratory testing or automation basics"* [S9]. Similarly, in a widely cited blog post, an engineer wrote: *"The questions were phrased to trick you into picking the textbook answer, not the one you'd use in practice"* [S5].

This concern is also reflected in critical analyses on testing blogs, which describe ISTQB as an academic exercise rather than a skills assessment. One article explained: *"You don't need practical experience to pass, just good memory and practice with sample questions"* [S2].

In summary, these voices portray ISTQB exams as relying too heavily on recall of theoretical constructs, which may undermine their ability to measure actual testing expertise. As a result, certified testers may carry the credential but lack the depth of applied skill expected in the workplace.

#### 5.2.6 Minimal Skill Development After Certification

Another recurring criticism is that obtaining an ISTQB certificate does not necessarily lead to improved practical testing skills. Practitioners stress that while candidates may gain theoretical knowledge, the certification does little to enhance their day-to-day effectiveness in real projects [S4, S6, S9, S12, S17]. This distinguishes the concern from critiques about exam design: here the issue is that even after passing, certified testers often show no noticeable change in ability.

On the LinkedIn platform, one QA lead wrote: *"We had team members pass the foundation exam, but in practice their test analysis and bug reporting did not improve at all"* [S9]. Similarly, in a Ministry of Testing discussion, a participant observed: *"The cert doesn't make you a better tester, it just means you memorized a syllabus"* [S12].

A blog post reinforced this perspective: *"I learned terms and models, but when I got back to work, nothing about how I approached testing really changed"* [S6]. Another practitioner reflected on Reddit: *"I passed the exam, but my exploratory testing was still weak – the certificate didn't fix that"* [S17].

Even managers express doubts about skill uplift. One remarked on LinkedIn: *"Certification may boost confidence, but it does not replace mentorship, practice, or project experience"* [S4]. Such comments highlight the perception that skills are built through real-world engagement rather than certification alone.

Collectively, these insights underline that ISTQB certification is often seen as providing knowledge recognition but not genuine skill advancement, raising questions about its long-term value for professional development.

#### 5.2.7 Context does matter: Regional and Contextual Limitations

Critics also emphasize that the value and recognition of ISTQB certification vary widely by region and organizational context [S2, S5, S11, S13, S18]. In some markets, particularly parts of Europe and Asia, the certification is viewed almost as a prerequisite, while in others, such as North America, it holds far less weight. This uneven recognition often shapes how testers perceive its utility.

On the Ministry of Testing platform, one practitioner remarked: *"In Germany and Switzerland, it's almost impossible to get a testing job without ISTQB – but in the US, nobody asks about it"* [S11]. Similarly, a blog post noted: *"In India, recruiters often filter resumes for ISTQB, but in Canada, hiring managers rarely care"* [S2].

Another LinkedIn discussion reinforced this disparity: *"The certification's importance depends on geography. In some regions, it's mandatory; in others, it's irrelevant"* [S18]. A tester writing on Testing Curator observed: *"ISTQB is treated like a gatekeeping credential in certain contexts, but that doesn't translate universally"* [S13].

This unevenness also extends across industry domains. For example, in heavily regulated sectors such as finance or healthcare, organizations may require certifications to satisfy compliance audits, whereas in fast-moving startups, employers prioritize demonstrable skills over certificates [S5].

Overall, these accounts highlight ISTQB as a context-dependent credential, whose perceived value is shaped less by universal acceptance and more by local hiring practices, cultural expectations, and industry-specific norms.

#### 5.2.8 Psychological and Cultural Effects

Some practitioners point to the psychological and cultural consequences of ISTQB certifications, arguing that the scheme fosters unhealthy perceptions of competence and status within testing teams [S4, S7, S11, S15, S18]. By creating a visible divide between "certified" and "non-certified" testers, ISTQB can unintentionally reinforce hierarchies and lead to misplaced confidence.

On the Ministry of Testing platform, one tester remarked: *"In my team, those with ISTQB thought they were automatically better testers – but in practice, some of the uncertified folks were far more skilled"* [S11]. A Reddit participant voiced a similar frustration: *"It creates this false prestige where people feel superior just because they have the cert"* [S15].

This effect is also noted on professional networks. A LinkedIn QA manager observed: *"Certification gave some testers confidence, but in a few cases it bordered on arrogance – they dismissed practices not in the syllabus"* [S18]. Another LinkedIn commenter added: *"I've seen new hires act entitled because of the certificate, which affected team dynamics negatively"* [S4].

In cultural terms, the certificate is sometimes treated as a badge of legitimacy, which may pressure newcomers to pursue it for acceptance rather than genuine learning. One user of the Ministry of Testing forum noted: *"It's not about what you know – it's about whether you tick the ISTQB box, otherwise you're not considered professional"* [S7].

These accounts highlight how ISTQB certification can generate psychological effects (overconfidence, insecurity, or entitlement) and cultural divides (us vs. them dynamics) that may disrupt team cohesion and professional development.

### 5.2.9 Over-standardization and de-facto Monopoly

A final strand of criticism concerns **the** dominant position of ISTQB in the certification landscape, which some practitioners describe as a de-facto monopoly that risks narrowing the field of testing knowledge [S2, S5, S9, S13, S18]. While having a widely recognized scheme brings consistency, many testers worry that over-standardization discourages alternative approaches and critical thinking.

On the Ministry of Testing platform, one participant remarked: *"ISTQB has become the default gatekeeper – if you don't have it, you're excluded, even if you have years of hands-on skill"* [S13]. Similarly, a LinkedIn practitioner observed: *"It dominates job ads and hiring practices to the point that other credible approaches are ignored"* [S18].

Some voices frame this as a risk to innovation in testing education. A Testing Curator article argued: *"When one framework becomes the de-facto standard, it can overshadow context-driven methods and exploratory testing schools of thought"* [S5]. Likewise, a blog post warned: *"We risk treating ISTQB as the only legitimate body of knowledge, which narrows diversity of perspectives in our craft"* [S2].

This concern extends to organizational adoption. A QA manager on LinkedIn wrote: *"Companies lean so heavily on ISTQB that they stop investing in internal training or other certifications"* [S9]. Such reliance can inadvertently create lock-in, where the testing community equates professionalization solely with ISTQB credentials.

Collectively, these perspectives highlight how ISTQB's **de-facto monopoly** status brings consistency but also the danger of over-standardization, potentially crowding out alternative, context-sensitive, and innovative views of what testing should be.

### 5.2.10 Criticism from Thought Leaders Undermining ISTQB's Authority

A further dimension of criticism comes from respected voices and thought leaders in the software testing community, who have openly questioned the credibility and usefulness of ISTQB certifications [S3, S11, S13, S19, S20]. Their views often carry significant weight, influencing practitioner sentiment and shaping industry discourse.

James Bach, a well-known figure in the context-driven testing movement, has consistently argued that ISTQB promotes a prescriptive and reductionist view of testing. In interviews and blog posts, he has described the certification as *"not a demonstration of skill but of compliance with a syllabus"* [S19]. Similarly, Michael Bolton has criticized the scheme as prioritizing terminology over critical thinking, remarking that *"ISTQB teaches you to parrot, not to test"* [S20].

On the Ministry of Testing platform, participants frequently cite these expert critiques, with one commenter noting: *"When leaders like Bach and Bolton dismiss ISTQB, it makes you wonder why we treat it as an authority"* [S11]. A Testing Curator article further amplified this sentiment: *"The strongest critics of ISTQB are also some of the most influential testers, which means their voices resonate widely"* [S13].

Even ISTQB-aligned sources acknowledge the challenge posed by these critiques. In the ASTQB FAQ, the organization itself highlights that while thought leaders disagree, ISTQB remains widely adopted, implicitly recognizing the tension between formal adoption and critical expert opinion [S3].

Taken together, these perspectives underscore how criticism from high-profile figures magnifies doubts about the certification's legitimacy. For many practitioners, if leading experts dismiss ISTQB as misaligned with real testing practice, it further undermines their confidence in its value.

### 5.3 Cross-Perspective Synthesis Between Endorsements and Criticisms (RQ3)

Although RQ1 and RQ2 present endorsements and criticisms separately for analytic clarity, several practitioner sources exhibited both positive and negative viewpoints within the same commentary. Because practitioner discussions in blogs, forums, and professional platforms are typically anonymous and fragmented, it was not objectively possible to quantify how many individuals simultaneously express both categories of viewpoints. However, we found that thematic cross-analysis revealed interesting co-occurrence patterns, as discussed next.

Practitioners often endorse ISTQB certifications for reasons such as:

- their signaling value in the job market,
- the shared terminology and structured vocabulary they provide, and
- their usefulness for onboarding junior testers.

The same practitioners may also criticize ISTQB certifications because of concerns related to:

- limited real-world applicability,
- exam-focused learning rather than skill development,
- outdated examples, or
- a perceived misalignment with modern testing practices.

These apparently contradictory viewpoints reflect the multifaceted role of ISTQB certifications. A certification can be valued for its organizational or career benefits while simultaneously being questioned for its pedagogical depth or its mapping to practical testing work. Therefore, the coexistence of endorsements and criticisms should be interpreted not as inconsistency but as evidence that ISTQB certifications serve different purposes for different practitioners.

This synthesis strengthens the connection between RQ1 and RQ2 by showing that practitioner perceptions are nuanced and multidimensional, rather than uniformly positive or negative.

## 6 Expert Review of Endorsements and Criticisms (RQ4 Results)

The third research question (RQ4) was addressed through a structured critical review conducted by a panel of four independent experts. Building on the practitioner dataset analyzed for RQ1 and RQ2, the experts were asked to evaluate the precision of endorsements and the fairness of criticisms. Our intent was that their judgments, gathered through two one-hour video meetings, would provide an external validation layer.

### 6.1 Approach for the critical review by the group of independent experts

The expert review was conducted in two online video meetings, each lasting one hour. We invited four experts, each bringing decades of experience in software testing and quality assurance across industry, consultancy, and academia.

To provide transparency on the expert panel, we ensured diversity across multiple dimensions. Two experts were drawn from academia (Europe and Asia) and two from industry (North America and Europe), reflecting a balance between research and applied perspectives. In terms of certification background, two panelists had extensive experience with multiple ISTQB certifications, while the other two had not pursued ISTQB qualifications, offering contrasting standpoints. We also confirmed that none of the panelists held formal roles within the ISTQB organization, reducing the risk of conflicts of interest. This diversity was intentional to capture a range of viewpoints across geographic, professional, and experiential contexts, thereby strengthening the credibility of the panel's judgments.

Prior to the meetings, experts were provided with the synthesized list of endorsements (RQ1) and criticisms (RQ2). They were asked to rate the precision of endorsements and the fairness of criticisms using a five-point Likert scale (very low → very high). This enabled a structured, quantitative baseline for comparison, which was then complemented by open discussion during the meetings.

The discussions were recorded and analyzed thematically. Beyond numerical ratings, the experts reflected on broader patterns, root causes behind criticisms, and differences between "schools of thought" in software testing; which we present and discuss in the next few subsections.

### 6.2 Precision of endorsements, as judged by the independent experts

The panel evaluated the precision of five categories of practitioner endorsements using a five-point Likert scale (1 = very low, 5 = very high).

Before presenting the expert evaluations, it is important to clarify how the constructs used in RQ4 were operationally defined for the panel. Experts were provided with explicit definitions to ensure consistent interpretation. Precision of endorsements referred to the extent to which each positive practitioner claim—derived from the MLR synthesis—was considered accurate, specific, and reflective of real-world testing practice. Experts judged whether an endorsement was articulated clearly, grounded in plausible practitioner experience, and free from overgeneralization. Fairness of criticisms referred to whether each negative practitioner statement represented a justified and proportionate concern. Experts assessed whether a criticism corresponded to credible challenges encountered in practice, avoided exaggeration, and acknowledged contextual variability. These definitions guided the expert panel in evaluating the practitioner-derived statements themselves, rather than the correctness of the MLR synthesis process.

Table 2 summarizes the ratings provided by the four experts. We should note that treating Likert scale categories as numeric values for calculating averages is debated in the research community [48], and we use the averages in Table 2 only as a pragmatic summary of expert judgments, not as precise measurements.

**Table 2-Precision of practitioner endorsements as judged by four independent experts**

| Endorsement item | Expert 1 | Expert 2 | Expert 3 | Expert 4 | Average |
|---|---|---|---|---|---|
| Career Advancement and Job Opportunities | 5 | 4 | 5 | 4 | 4.5 |
| Professional Recognition and Credibility | 4 | 3 | 4 | 4 | 3.75 |
| Foundational Knowledge and Common Terminology | 5 | 5 | 4 | 5 | 4.75 |
| Structured Approach and Testing Efficiency | 3 | 4 | 2 | 3 | 3.0 |
| Organizational and Global Advantages | 4 | 3 | 3 | 4 | 3.5 |

The results show the general agreement among the experts in the precision of endorsements, synthesized from the online resources of practitioners at large. Foundational Knowledge and Common Terminology received the highest average rating (4.75), confirming that ISTQB has successfully standardized language and concepts across the profession. Career Advancement and Job Opportunities were also rated highly (4.5), reflecting strong agreement that certification is widely recognized by employers, especially at entry level.

Endorsements around Professional Recognition and Credibility (average response=3.75 out of 5) and Organizational and Global Advantages (average response=3.5 out of 5) were judged moderately precise. Experts noted that recognition often varies by region and industry, and that organizational benefits are sometimes overstated in practitioner narratives.

Finally, Structured Approach and Testing Efficiency received the lowest precision rating (average response=3.0 out of 5). While some experts acknowledged the value of a structured framework, others cautioned that ISTQB syllabi do not automatically lead to efficiency gains in practice.

### 6.3 Fairness of criticisms, as judged by independent experts

The panel assessed the fairness of ten criticisms derived from practitioner sources. While opinions expressed on forums and blogs are inherently informal, they remain valuable indicators of practitioner sentiment. Their credibility must therefore be considered through structured evaluation. Table 3 presents the fairness ratings given by the four experts using a five-point Likert scale (1 = very unfair, 5 = very fair).

**Table 3-Fairness of practitioner criticisms as judged by four independent experts**

| Criticism item | Expert 1 | Expert 2 | Expert 3 | Expert 4 | Average |
|---|---|---|---|---|---|
| Primarily theoretical / not practically useful | 4 | 5 | 4 | 4 | 4.25 |
| Not a proper measure of testing skill | 5 | 4 | 4 | 4 | 4.25 |
| Overemphasis on memorization | 5 | 4 | 5 | 4 | 4.50 |
| Outdated content / slow adaptation | 3 | 4 | 3 | 3 | 3.25 |
| Minimal skill development after certification | 3 | 3 | 4 | 3 | 3.25 |
| Commercialization / financial motives | 2 | 3 | 3 | 4 | 3.00 |
| Regional and contextual limitations | 3 | 4 | 4 | 3 | 3.50 |
| Over-standardization / de-facto monopoly | 3 | 2 | 3 | 4 | 3.00 |
| Psychological and cultural effects | 2 | 2 | 3 | 2 | 2.25 |
| Criticism from thought leaders undermining ISTQB | 3 | 3 | 4 | 2 | 3.00 |

The experts agreed that the criticisms are not simple "true or false" statements. They are complex issues whose validity depends heavily on context. For the exam-related items, such as overemphasis on memorization (avg 4.50), primarily theoretical (4.25), and not a proper measure of testing skill (4.25), there was strong consensus that these are fair criticisms across most situations, since multiple-choice formats cannot fully capture practical competence.

Other criticisms, like outdated content (3.25), minimal skill development (3.25), and regional and contextual limitations (3.50), were rated in the middle range. Here, experts stressed *"it depends"*: the criticism may be valid in certain geographies, organizational cultures, or earlier syllabus versions, but less so where updates and adaptations have already taken place.

For criticisms about the wider ecosystem and perceptions, such as commercialization (3.00), over-standardization (3.00), psychological and cultural effects (2.25), and influence of thought leaders (3.00), the panel again emphasized context. These may resonate strongly with some practitioners but not with others. For instance, commercialization may be a real concern where training costs are high, but less relevant in markets with affordable local training. Likewise, cultural effects can be significant in certain workplace environments but may not apply universally.

In short, the panel's conclusion was that every criticism has a degree of truth, but its fairness depends on the context in which it is raised. This reflects a recurring theme in SE: most debates ultimately lead back to *"it depends."*

## 6.4 General root causes for the criticisms, as discussed by the independent experts

Beyond rating individual endorsements and criticisms, the experts also reflected on the **underlying reasons** why these criticisms keep surfacing in practitioner discussions. Their view was that most concerns are not isolated, but trace back to deeper structural issues in the testing community worldwide, which is indeed very large and diverse.

The panel identified two such general root causes that we discuss next.

### 6.4.1 Different "schools of thought" in testing

One general root cause identified by the panel was the existence **of** different "schools of thought" [49] in software testing, each carrying its own philosophy **of** what matters in the discipline. This diversity of viewpoints explains why endorsements and criticisms of ISTQB certifications can appear contradictory yet still be valid.

The experts noted that the Analytical school values systematic, knowledge-based approaches, where ISTQB fits well by providing a structured body of concepts and terminology. The QA and Factory schools likewise see value in standardization, documentation, and globally recognized certification frameworks. By contrast, the Context-Driven and Agile schools of testing emphasize situational adaptability, hands-on skill, and critical thinking; from this perspective, ISTQB's exam-driven model is often criticized as too theoretical or detached from practice.

The tension between these schools means that criticisms such as "too theoretical" or "not a measure of real skill" are valid for practitioners working in context-driven or agile environments**,** but less so for those in organizations that prioritize standardized processes and shared terminology. Conversely, endorsements around career recognition and common language are highly valued in structured, process-heavy contexts, while being downplayed in more exploratory settings.

The panel concluded that these contrasting schools of thought form a major root cause of polarized opinions on ISTQB. In software engineering—as in many other fields—the answer to whether a certification is valuable or not often comes back to the familiar conclusion: *"it depends."*

### 6.4.2 One size does not fit all, and impact of context

The second root cause identified by the panel is the reality that no single certification framework can meet the needs of all practitioners, organizations, and regions. ISTQB syllabi are designed to be globally applicable, but this broad scope inevitably leads to gaps when applied in specific contexts.

For example, in some markets, ISTQB serves as **a** baseline entry requirement**, providing** recognition and career mobility. In others, especially where testing is more practice-oriented, its emphasis on structured knowledge is perceived as **too theoretical**. Likewise, organizations with mature, process-driven QA cultures often see strong value in standardized terminology, while agile or start-up environments find less relevance.

The experts highlighted that these contextual differences explain why the same criticism may resonate strongly with one group but not with another. For instance, *"minimal skill development after certification"* may hold true for experienced testers expecting advanced practice, but is less applicable for newcomers who primarily need shared vocabulary and foundational concepts. Similarly, *"commercialization"* can be a pressing issue in regions where training costs are high, yet hardly noted in places where affordable community-led training exists.

In essence, the panel stressed that one size does not fit all. The fairness of each criticism depends on factors such as career stage, organizational culture, local hiring practices, and even regional economics. This context-dependency reinforces a broader truth in software engineering debates: the validity of a claim often comes down to *"it depends."*

## 6.5 Recent ISTQB Initiatives Addressing Criticisms

The panel noted signs of responsiveness from ISTQB. Two directions stood out: (1) engaging with practice-centric credentials in test automation, and (2) introducing new certifications aligned with fast-moving topics.

### 6.5.1 Partnering with other certifications focusing on test automation

One noteworthy initiative has been ISTQB's decision to endorse the Practical Tester certification by A4Q as an "add-on" certification (see: istqb.org/istqb-endorses-the-practical-tester-by-a4q). This move directly addresses one of the long-standing criticisms of ISTQB: that its exams largely emphasize theoretical knowledge over hands-on ability. The A4Q Practical Tester is explicitly designed to validate practical skills in applying testing techniques. By endorsing this credential, ISTQB acknowledges that effective tester competence involves not only understanding the syllabus but also demonstrating applied skills in realistic testing scenarios.

The panel viewed this as a significant step, since it signals that ISTQB is opening up its certification ecosystem to complementary initiatives that provide stronger evidence of practice. However, the experts also noted that the effectiveness of this collaboration will depend on how well the integration is communicated to employers and whether it truly influences hiring and promotion decisions.

### 6.5.2 Developing New Certifications Aligned with Industry Trends

Another important initiative highlighted by the panel is ISTQB's expansion into Artificial Intelligence (AI)–related certifications, designed to address the repeated criticism that ISTQB syllabi lag behind industry trends. Two recent certifications stand out.

The first is the Certified Tester – Testing with (using) Generative AI (CT-GenAI)[1]. This certification focuses on how testers can responsibly apply generative AI tools within their daily activities. It covers practical issues such as test case generation, test data creation, and the evaluation of AI-produced artifacts, while also addressing risks like bias, hallucinations, and trustworthiness. By formalizing these topics, ISTQB acknowledges the growing role of generative AI in software testing and provides a structured framework for practitioners to engage with these tools.

The second is the Certified Tester AI Testing (CT-AI)[2]. Unlike CT-GenAI, which is about using AI in testing, CT-AI emphasizes the reverse: testing AI systems themselves. The syllabus covers fundamental AI/ML concepts, risk categories for AI systems, and appropriate test strategies to evaluate quality attributes such as robustness, fairness, and explainability. This certification responds directly to the demand from industry for testers capable of validating AI-driven systems, a domain where existing ISTQB syllabi had little coverage.

# 7 DISCUSSIONS, IMPLICATIONS AND LIMITATIONS

In this section, we step back to discuss the broader meaning of these findings. We consider their implications for different stakeholders—ISTQB itself, testers, employers, and researchers—while also reflecting on the lessons learned from using AI to support the MLR. We will also outline limitations and validity threats that should be borne in mind when interpreting the study's contributions.

A key methodological consideration concerns how the AI-derived practitioner insights and expert panel evaluations were integrated. The AI-assisted MLR offered substantial breadth by surfacing diverse practitioner viewpoints at scale, whereas the expert panel contributed depth, contextual interpretation, and assessments of practical validity. Because AI tools cannot reliably evaluate the soundness, fairness, or contextual appropriateness of practitioner statements, cross-verification (RQ3) required human analytical judgment rather than AI automation alone. The synthesis presented in Section 7 therefore reflects an intentional triangulation: endorsements and criticisms were first identified through AI-assisted analysis, then interpreted through expert reasoning. This combination enhances efficiency while also strengthening trustworthiness, as no single evidence source was treated as definitive.

---

[1] istqb.org/certifications/gen-ai/

[2] istqb.org/certifications/certified-tester-ai-testing-ct-ai

## 7.1 Implications, recommendations and suggestions

### 7.1.1 Implications for ISTQB certifications

The analysis shows that while ISTQB holds global recognition, persistent criticisms highlight structural gaps. Four areas merit attention: (1) Theory-practice balance: Exams should incorporate performance-based elements (scenario analysis, defect detection) to address memorization concerns. (2) Transparency: Clearer separation between certification governance and the commercial training market would mitigate commercialization perceptions. (3) Syllabus agility: Shorter update cycles across all syllabi and closer engagement with industry consortia are essential to keep pace with emerging practices. (4) Contextual enrichment: Training materials could incorporate domain-specific case studies and regional examples without diluting global standardization.

### 7.1.2 Implications for practitioners and organizations

For testers, ISTQB remains valuable as a baseline credential for entry and career mobility, but should be complemented with practical demonstrations of competence (bug bounties, open-source projects, performance-based platforms). Testers should view certification as a starting point, weighing its cost against alternative learning opportunities.

For employers, ISTQB provides a useful common language and entry-level filter but is not a reliable proxy for practical skill. Certifications should be combined with performance-based assessments, practical interviews, and coding exercises. For specialist or senior roles, hands-on experience and domain-specific skills should take precedence.

### 7.1.3 Implications for academia and research

Universities should use ISTQB syllabi as a reference point while maintaining strength in hands-on projects with industry-strength tools and exposure to diverse schools of thought in testing. This complementary approach ensures graduates master both foundational frameworks and the complexity of real-world practice.

Researchers should conduct rigorous empirical studies to move beyond opinion-driven debates, investigate alternatives to certification (serious games, performance-based assessments), and bridge the academia-industry gap through evidence-based curriculum alignment.

## 7.2 Experience and insights regarding the AI-assisted MLR work

This study also provides reflections on the process of conducting an AI-assisted Multivocal Literature Review (MLR). The experience showed that AI can serve as a powerful assistant in handling heterogeneous sources, ranging from academic papers to grey literature such as blogs, forums, and organizational reports. By treating these sources as a qualitative dataset, the AI facilitated systematic coding, clustering of endorsements and criticisms, and synthesis of themes.

One insight was the importance of carefully crafted prompts and human oversight. The AI was able to conduct thematic analysis, but intermediate outputs required inspection to prevent misclassification or oversimplification. This aligns with the broader lesson that AI can accelerate evidence synthesis, but it cannot replace the role of researchers in ensuring accuracy and contextual interpretation.

Another lesson concerned transparency and reproducibility. Documenting the instructions given to the AI and the subsequent human validation steps was essential to establish credibility. Without such transparency, AI-assisted reviews risk being perceived as opaque or biased.

Finally, the work highlighted the potential of AI tools to scale evidence-based software engineering research. Traditional MLRs are time-consuming and resource-intensive; AI support offers a way to accelerate synthesis while still maintaining rigor, provided that proper validation safeguards are in place.

In sum, our experience suggests that AI-assisted MLRs are a promising complement to conventional reviews, enabling researchers to handle larger and more diverse evidence bases. However, their effectiveness depends on human–AI collaboration, with AI providing speed and breadth, and researchers ensuring depth, accuracy, and trustworthiness.

## 7.3 Alternatives to ISTQB certifications

As discussed in Section 2.4m, there are many certifications in the area of software testing, which would be the natural alternatives to ISTQB certifications.

We discuss in the following two other alternatives: (1) platforms for assessment of job applicants, and (2) serious-game–based tools.

An online platform named CatchyApp by a firm named CatchyLabs (CatchyLabs.co) positions itself as "the new norm for quality assurance hiring." Instead of relying on certificates as proxies for competence, CatchyApp enables candidates to participate in structured, practical interviews where they perform testing-related tasks. Hiring managers receive evidence-based insights into a candidate's actual abilities, reducing reliance on gut feelings or theoretical credentials. This approach speaks directly to criticisms of ISTQB certifications as insufficient indicators of real skills, offering organizations a way to evaluate testers based on performance rather than formal exam results.

The second example is BugHunt [50], a serious-game–based tool developed for teaching and practicing defect-finding skills. BugHunt presents learners with interactive software environments in which they must locate bugs, making early software testing education more engaging and practical. Research shows that such game-based learning improves motivation and fosters deeper understanding of testing concepts. While not a certification per se, BugHunt illustrates how applied, experiential tools can substitute or complement formal pathways in skill development.

We believe that these alternatives highlight a growing shift: competence in testing is increasingly validated by what practitioners can do, not just by what they know. While ISTQB retains its global recognition and standardization role, platforms like CatchyLabs and BugHunt demonstrate that organizations are exploring more direct, practice-based ways to assess and develop tester competence.

## 7.4 Limitations and Validity Threats

We discuss in this section the potential threats to validity of the study and the actions we took to minimize or mitigate them, based on the guidelines in software engineering [51].

Given the nature of our study, it is important to discuss limitations and validity threats from two broad angles: (1) validity threats for the study's core contribution: synthesis of practitioners' endorsements and criticisms; (2) validity threats for the AI-assisted MLR (process).

In our discussions below, we categorize threats into the main four validity aspects [51]: internal validity, external validity, construct validity, and conclusion validity.

### 7.4.1 Validity threats in synthesizing practitioner perspectives

The core contribution of this study lies in the synthesis of practitioner endorsements and criticisms of ISTQB certifications. While the process followed MLR guidelines, several validity threats must be acknowledged.

Internal validity. Practitioner statements collected from blogs, forums, and online articles may reflect individual experiences, biases, or local contexts. To mitigate this, we included a diverse set of 20 sources covering different roles and geographies, and we triangulated them against existing literature and recent ISTQB initiatives.

External validity. The dataset cannot claim to represent the views of all testers worldwide. Our findings therefore capture a sample of practitioner voices rather than the full population. To reduce this threat, we selected sources from multiple countries and roles (testers, managers, trainers), ensuring breadth and diversity.

Construct validity. Another threat lies in the classification of statements into "endorsements" and "criticisms." Ambiguities exist where the same source expresses both positive and negative sentiments. To address this, we applied thematic coding carefully, separating mixed statements and assigning them to both categories where appropriate.

Conclusion validity. A final threat concerns the strength of conclusions drawn. Practitioner perceptions are inherently context-dependent, and as discussed in Section 8.4, many criticisms reduce to "it depends." Our conclusions should therefore be interpreted as patterns of perception, not as universally generalizable truths.

In summary, while the synthesis provides valuable insights into how ISTQB certifications are endorsed and critiqued in practice, the results must be viewed within the limitations of qualitative evidence. The safeguards applied—diversity of sources, careful coding, and explicit acknowledgment of context-dependency—help to mitigate these threats but do not eliminate them entirely.

### 7.4.2 Validity threats for the AI-assisted MLR (process)

The second set of validity concerns relates to the AI-assisted MLR process itself. While the use of AI enabled efficient handling of a diverse dataset, it also introduces unique risks.

**Internal validity.** AI models can misinterpret practitioner statements, oversimplify nuanced arguments, or misclassify sentiments. To reduce this threat, all intermediate outputs were inspected and corrected by the researchers. Representative quotes were always cross-checked with their original sources.

**External validity.** Since AI-generated coding decisions depend on the model and the prompts used, reproducibility may vary if different AI systems or instructions are applied. To mitigate this, we documented our prompting strategy and analysis steps in detail, enabling other researchers to replicate or extend the process.

**Construct validity.** There is a risk that the AI's thematic clustering may not fully reflect the intended constructs of "endorsement" or "criticism." We addressed this by treating AI outputs as **initial suggestions** rather than final categorizations, and by applying human judgment in resolving ambiguities.

**Conclusion validity.** Another concern is the averaging and summarization of expert ratings and AI-coded themes. Treating Likert-scale categories as numeric averages, for instance, is debated in the research community. We applied averages only as pragmatic summaries, interpreting them cautiously as indicators of relative patterns rather than precise measurements.

In summary, the AI-assisted process enhanced the efficiency and scale of the review, but the safeguards of **human** oversight**,** transparent documentation, and careful interpretation were essential to ensure credibility. The results should thus be seen as the outcome of a human–AI collaboration, where AI provided speed and breadth, and researchers ensured rigor and validity.

## 8 CONCLUSIONS AND FUTURE WORKS

This study has provided a critical reflection on ISTQB certifications by synthesizing practitioner endorsements and criticisms through an AI-assisted Multivocal Literature Review, and by complementing those insights with an expert review of their precision and fairness. The results highlight a nuanced picture: ISTQB offers global recognition, shared terminology, and a structured foundation, but persistent concerns remain about its theoretical orientation, memorization-heavy exams, and limited measurement of practical skills.

We also discussed broader implications for stakeholders. For ISTQB itself, the findings suggest the need for more practice-relevant and agile syllabi. For testers, certifications can open career doors but should be complemented with demonstrated competence. For employers, ISTQB provides a useful signal of baseline knowledge but must be combined with performance-based assessments. For universities, ISTQB syllabi can serve as reference points but should be enriched with real-world projects, industry-strength tools, and critical perspectives. For researchers, the study underlines the importance of more empirical work on certifications and of exploring alternatives such as serious games and performance-based platforms.

Beyond substantive findings, this study contributes methodologically by demonstrating the potential of AI-assisted MLRs. AI enabled the systematic handling of diverse practitioner voices, but the process also highlighted the importance of human oversight, transparent documentation, and careful interpretation.

Future work can proceed in several directions. First, broader empirical studies, including large-scale surveys and interviews, are needed to deepen and validate the patterns observed here. Second, comparative research across different testing certifications and alternative competence-assessment platforms could provide richer insights into the evolving landscape of professional qualifications. Third, further exploration of AI-supported methods for evidence synthesis could help scale up multivocal reviews while maintaining rigor.